\documentclass[aps,pre,twocolumn,superscriptaddress]{revtex4-1}

\usepackage[utf8]{inputenc}
\usepackage[T1]{fontenc}
\usepackage{lmodern}
\usepackage{amsmath,amsfonts,amssymb}
\usepackage{graphicx}

\usepackage{hyperref}
\usepackage[usenames,dvipsnames]{xcolor}
\hypersetup{colorlinks=true, linkcolor=blue!50!black, urlcolor=blue!50!black, citecolor=blue!50!black}
\usepackage[all]{hypcap}

\usepackage[arrow, matrix, curve]{xy}
\usepackage{bm}
\usepackage{yhmath}
\usepackage{color}
\usepackage[normalem]{ulem} 
\newcommand\diff{\mathrm{d}}

\usepackage[normalem]{ulem}

\renewcommand{\vec}[1]{\mathbf{#1}}
\renewcommand{\imath}[0]{\mathsf{i}}

\begin{document}

\title{Nonergodicity parameters of confined hard-sphere glasses}

\author{Suvendu Mandal}
\author{Simon Lang}
\affiliation{Institut f\"ur Theoretische Physik, Universit\"at Innsbruck, Technikerstr. 21A, A-6020 Innsbruck, Austria}

\author{Vitalie Bo\ifmmode \mbox{\c{t}}\else \c{t}\fi{}an} 
\affiliation{Lehrstuhl f\"ur Technische Thermodynamik, RWTH Aachen University, D-52062 Aachen, Germany}

\author{Thomas Franosch}
\email{Thomas.Franosch@uibk.ac.at}
\affiliation{Institut f\"ur Theoretische Physik, Universit\"at Innsbruck, Technikerstr. 21A, A-6020 Innsbruck, Austria}

\begin{abstract}
Within a recently developed mode-coupling theory for fluids confined to a slit we elaborate numerical results for  the long-time limits of suitably generalized intermediate scattering functions. The theory requires as input  the density profile perpendicular to the plates, which we obtain from density functional theory within the fundamental-measure framework, as well as  symmetry-adapted static structure factors which can be  calculated relying on the inhomogeneous Percus-Yevick closure. Our calculations for the  nonergodicity parameters for both the collective as well as for the self motion are in qualitative agreement with our extensive event-driven molecular dynamics simulations for the intermediate scattering functions for slightly polydisperse hard-sphere systems at high packing fraction. We show that the variation of the nonergodicity parameters as a function of the wavenumber correlates with the  in-plane static structure factors, while subtle effects become apparent in the structure factors and relaxation times of higher mode-indices. A criterion to predict the multiple reentrant from the variation of the in-plane static structure is presented.
\end{abstract}

\maketitle

\section{Introduction}
The structural relaxation of dense liquids exceeds microscopic time scales by orders of magnitude upon approaching the glass transition via compression or cooling the system.
The origin of the glass transition remains heavily debated~\cite{Adam1965,Pusey1987,Stillinger1995,Ediger1996, Debenedetti2001,Cipelletti2005,Biroli2006,Heuer2008,Candelier2010,Berthier2011a,Berthier2011b,Hunter2012,Sengupta2012,Martinez2013} 
and still constitutes a challenge for theory, simulation, and laboratory experiments, 
although many facets of the phenomena associated with the glass transition have been rationalized successfully within the mode-coupling theory of the glass transition~\cite{Gotze_Complex_Dynamics,Voigtmann2009,Voigtmann2011,Sperl2010,Gnan2014}  developed during the last thirty years. 

Microscopically, particles become trapped by transient cages comprised of their surrounding neighbors, suggesting that the local structure plays a predominant role for the drastic slowing down of the structural relaxation. 
Correspondingly, one anticipates that nanoconfinement and wall-particle interactions introduce competing mechanisms 
strongly affecting their structural as well as their transport properties~\cite{Lowen2001,Alba-Simionesco2006}. In fact, these competing mechanisms are expected to be very strong when the wall-to-wall separation is of a few particle diameters only. 

Confinement is of considerable interest also in a variety of physical, chemical, and biological systems~\cite{Zhou2008}. For example, such strong confinements naturally occur in porous rocks and biological crowded systems such as living cells~\cite{Zhou2008}. A molecular-level understanding of such confinement effects is essential to design coatings, nanopatterning, and fabrication of nanomaterials~\cite{Cipelletti2005,Mattsson2009}.

Recently, confinement effects for glass-forming liquids have been investigated  in slit geometry
by computer simulations~\cite{Scheidler2000,Scheidler2000b,Scheidler2002,Scheidler2004,
Varnik2002c,Varnik2002d,Torres2000,Baschnagel2005,Mittal2006,Mittal2007,Mittal2007b,Mittal2008,Krishnan2003, Krishnan2012,Goel2008,
Krekelberg2011, Ingebrigtsen2013, Ingebrigtsen2014,Saw2016} as well as 
laboratory 
experiments~\cite{Nugent2007,Sarangapani2011,Sarangapani2012,Hunter2014,Williams2015,Nygard2012,Nygard2013,Nygard2016a,Nygard2016b,Kienle2016,Zhang2016,Ghosh2016} focusing on the regime of moderate confinement with slit widths of several particle diameters or larger. These studies demonstrate how confinement affects the dynamics of dense liquids for various particle-wall interactions or wall roughnesses. The dynamics in confinement has been shown to  increase or decrease compared to the bulk depending in a subtle way on the roughness of the walls~\cite{Baschnagel2005,Krekelberg2011}. However, the question what controls the dynamics of inhomogeneous liquids in confinement has remained elusive so far. The role of local order is emphasized within a remarkable empirical scaling of the diffusivities or structural relaxation times with the excess entropy~\cite{Mittal2006,Mittal2007,Mittal2008,Ingebrigtsen2013}.

A complementary microscopic approach is provided by the mode-coupling theory (MCT)~\cite{Gotze_Complex_Dynamics} which predicts a two-step relaxation for bulk liquids close to the glass transition accompanied by a series of scaling laws. It has been shown by computer simulations that many of the features of the MCT persist even in porous confinements~\cite{Gallo2000a, Gallo2009, Gallo2012}. An extension of the MCT to frozen disordered host structures has been developed~\cite{Krakoviack2005,Krakoviack2007, Krakoviack2009, Krakoviack2011,Szamel2013} predicting a subtle reentrant phenomenon as the fraction of liquid particles in the system is varied. Parts of the predictions have been verified also in simulations~\cite{Kurzidim2009,Kurzidim2010,Kurzidim2011, Kim2009,Kim2011}. 
In these porous media spatial correlation functions are isotropic and translationally invariant after averaging over different realizations of the disorder. 

In contrast, dense liquids squeezed into a narrow channel display an inhomogeneous density profile in the direction perpendicular to the walls. 
Recently, the MCT has been extended also to describe dense liquids in such planar confinements~\cite{Lang2010,Lang2012} relying on symmetry-adapted modes that account for  the broken translational symmetry perpendicular to the walls. The theory displays unique solutions which reflect all properties of correlation functions~\cite{Lang2013b} and reproduces the limits of a bulk system as well as of  a two-dimensional liquid as the wall separations
 becomes large or small~\cite{Lang2014b}. Surprisingly, for small wall separation the lateral and transverse degrees of freedom decouple~\cite{Franosch2012,Lang2014c} and a slow divergent time scale emerges controlling the crossover from 2D to 3D systems~\cite{Mandal2017a}.
 The tagged-particle dynamics for slit geometry has also been elaborated within MCT~\cite{Lang2014a}.

A striking prediction of the MCT in slit geometry has been the emergence of a multiple reentrant glass transition in the nonequilibrium state diagram as a function of the slit width along lines of constant packing fractions~\cite{Lang2010}. This scenario has been corroborated by event-driven~\cite{Alder1957,Rapaport1980,Bannerman2011} molecular dynamics simulations for slightly polydisperse hard-sphere systems~\cite{Mandal2014b, Varnik2016} upon measuring the self-diffusion coefficients parallel to the walls and extrapolating isodiffusivity lines to the glass-transition line.
 The multiple reentrance is attributed  to a complex competition between the layering induced by the walls and local caging.

Although the MCT captures the overall behavior of the nonequilibrium state diagram, many aspects associated with the dynamics in confinement have not been worked out so far.  For example, the matrix-valued character of the static structure factors has not been tested explicitly by experiments or computer simulations. The MCT in confinement requires structural quantities as input, hence a comparison of liquid state theory with simulation results is highly desirable. Furthermore, the theory allows calculating the intermediate scattering functions which are measurable  quantities in computer simulations or experiments. Then one would like to know how the associated nonergodicity parameters, i.e. the plateau values at intermediate time scales, behave as a function of the wavenumber and how they correlate with the generalized static structure factors.

The goal of the present paper is to further elaborate on the glassy dynamics in confinement and to provide a comparison between numerical results of the MCT in confinement to event-driven molecular dynamics simulations for slightly polydisperse hard-sphere systems. 
We compare simulations for the density profile to numerical results obtained from density-functional theory with fundamental-measure functionals~\cite{Roth2010,Hansen2006} and  the matrix-valued static structure factors  including now higher-order modes obtained using the inhomogeneous Percus-Yevick closure~\cite{Nygard2012,Nygard2013, Hansen:Theory_of_Simple_Liquids,Henderson:Fundamentals_of_inhomogeneous_fluids,Ram2014}. Then we discuss for the first time the nonergodicity parameters from the  MCT equations in  the long-time limit for the coherent dynamics. We also present new results for the tagged-particle motion. The nonergodicity parameters will be discussed as a function of wavenumber and compared to our new  simulation results for the time-dependent intermediate scattering functions covering the full range of wavenumbers.   
In particular, we identify key structural features that are responsible for the non-monotonic behavior in the phase diagram.

\section{Mode-coupling theory} \label{sec:simulation-setup}
\label{sec:theory}

Here we  introduce  the notation for the relevant quantities and 
provide a summary of the mode-coupling equations in confined geometry, for a detailed derivation of these equations see Ref.~\cite{Lang2010,Lang2012}.
The theory considers a single-component liquid comprised of  $N$ identical structureless particles of mass $m$ confined between two plane parallel smooth hard walls. Particle positions and momenta are specified by  $\vec{x}_n=(\vec{r}_n, z_n)$ and $\vec{p}_n=(\vec{P}_n, P_n^z)$ for $n=1,\ldots, N$, where $\vec{r}_n$ and $\vec P_n$ describe the  in-plane coordinates and momenta, respectively. The confinement restricts the transverse positions to $-L/2 \leq z_n \leq L/2$. For particles with hard-core repulsion of exclusion radius $\sigma/2$, the physical wall separation then corresponds to $H=L+ \sigma$. 

The confinement renders the liquid non-uniform in the direction perpendicular to the walls, in particular, it induces a modulation of the equilibrium density profile $n(z)$. The associated Fourier components are obtained as
\begin{equation}\label{eq:density}
  n_\mu = \int \diff z \exp( \text{i} Q_\mu z) n(z),
\end{equation}
where the mode index $\mu \in \mathbb{Z}$ is  discrete and corresponds to wavenumbers $Q_{\mu}=2\pi \mu /L$. Here integrals 
for transverse degrees of freedom are restricted to the accessible slit width, $z\in [-L/2,L/2]$. In particular, $n_0$ corresponds to the average particle number per area. A similar decomposition into Fourier modes $v_\mu$ holds for the local volume $v(z) = 1/n(z)$. 

Symmetry-adapted microscopic fluctuating density modes are introduced
\begin{equation}
\rho_{\mu}(\vec{q},t)=\sum_{n=1}^{N} \exp[\text{i} Q_\mu z_n (t)] \, \text{e}^{\text{i} \vec{q} \cdot \vec{r}_n(t)},
\end{equation}
where $\vec{q}=(q_{x},q_{y})$ are the conventional (discrete for finite cross sectional area $A = L_\text{box}^2$) wavevectors in the lateral 
plane. The key quantity in our discussion will be the associated coherent time-dependent correlation function
\begin{equation}\label{eq:scattering_function}
S_{\mu\nu}(q,t)=\frac{1}{N} \langle\rho_{\mu}(\vec{q},t)^* \rho_{\nu}(\vec{q},0)  \rangle,
\end{equation}
also referred to as the generalized intermediate scattering function, 
which measures the correlated particle motion over time $t$ and at inverse length scale $q$.
 Its initial value $S_{\mu\nu}(q) = S_{\mu\nu}(q,t=0)$ is the proper generalization of the static structure factor to slit geometry. 
The dependence on the wavenumber $q$ is suppressed  here and in the following if all quantities in an equation refer to the same wavenumber.  Similarly if the time variable is not displayed explicitly, it refers to time zero. 
The connection to the corresponding direct correlation function $c_{\mu\nu}(q)$ is provided by the Ornstein-Zernike relation~\cite{Hansen:Theory_of_Simple_Liquids,Henderson:Fundamentals_of_inhomogeneous_fluids}, which reads upon decomposition into symmetry-adapted modes~\cite{Lang2010,Lang2012}
\begin{equation}\label{eq:OZ_equation}
 \mathbf{S}^{-1} = \frac{n_0}{L^2} [ \mathbf{v} - \mathbf{c} ] .
\end{equation}
  Here bold symbols refer to matrices in the mode-indices, $[\mathbf{S}]_{\mu\nu} = S_{\mu\nu}$, and a natural matrix notation has been employed. 
The matrix corresponding to the local volume is provided by $[\mathbf{v}]_{\mu\nu}  = v_{\nu-\mu}$.

Using the Zwanzig-Mori projection operator formalism~\cite{Gotze_Complex_Dynamics,Hansen:Theory_of_Simple_Liquids}
exact equations of motion for the  collective correlators $S_{\mu \nu}(q,t)$ have been derived~\cite{Lang2010,Lang2012} to 
\begin{equation}\label{eq:eom1}
 \dot{\mathbf{S}}(t) + \int_0^t \mathbf{K}(t-t') \mathbf{S}^{-1} \mathbf{S}(t') \diff t' = 0.
\end{equation}
A non-trivial feature of the theory is that the current kernel $\mathbf{K}(t)$ naturally splits into decay channels parallel and perpendicular to the walls
\begin{equation}\label{eq:split}
 {K}_{\mu\nu}(q,t) = \sum_{\alpha\beta=\parallel,\perp}b^{\alpha}(q,Q_{\mu}){\mathcal{K}}^{\alpha\beta}_{\mu\nu}(q,t)
 b^{\beta}(q,Q_{\nu}).
\end{equation}
with channel index $\alpha=\parallel,\perp$. Here the selector 
$b^{\alpha}(x,z)=x \delta_{\alpha\parallel}+z\delta_{\alpha\perp}$ has been introduced for a compact notation. Caligraphic symbols are used for quantities associated with both a channel index as well as a mode index, i.e. for the channel-resolved current $[\boldsymbol{\mathcal K}(q,t) ]^{\alpha\beta}_{\mu\nu} = {\mathcal K}^{\alpha\beta}_{\mu\nu}(q,t)$, and again a natural matrix notation will be used. 

A second Zwanzig-Mori projection step for the case of Newtonian dynamics yields another exact equation of motion
\begin{equation}\label{eq:eom_Newton}
\boldsymbol{\mathcal J}^{-1} \dot{\boldsymbol{\mathcal K }}(t) + \int_0^t  \boldsymbol{\mathcal M}(t-t') \boldsymbol{\mathcal K}(t') \diff t' = 0 ,
\end{equation}
with $\boldsymbol{\mathcal J} = \boldsymbol{\mathcal K}(t=0)$ and the force kernel $\boldsymbol{\mathcal M}(t)$. 

The mode-coupling ansatz~\cite{Gotze_Complex_Dynamics} prescribes a strategy to approximate the force kernel in terms of a bilinear functional in the intermediate scattering functions itself. For the case of slit confinement the MCT yields~\cite{Lang2010,Lang2012} 
\begin{align}\label{eq:MCT_ansatz}
 {\mathcal M}_{\mu\nu}^{\alpha\beta}(q,t) =& {\mathcal F}_{\mu\nu}^{\alpha\beta}[\boldsymbol{S}(t);q] \nonumber \\
=& \frac{1}{2N} \sum_{\vec{q}_1,\vec{q}_2 = \vec{q}-\vec{q}_1}  \sum_{\mu_{1}\mu_{2} \atop \nu_{1} \nu_{2}} \mathcal{Y}^\alpha_{\mu,\mu_{1} \mu_{2}}(\vec{q},\vec{q}_{1}\vec{q}_{2}) \nonumber \\
&\times  S_{\mu_{1}\nu_{1}}(q_{1},t) S_{\mu_{2}\nu_{2}}(q_{2},t)  \mathcal{Y}^\beta_{\nu,\nu_{1} \nu_{2}}(\vec{q},\vec{q}_{1}\vec{q}_{2})^*, 
\end{align}
where the  vertices $\mathcal{Y}^\alpha_{\mu,\mu_{1} \mu_{2}}(\vec{q},\vec{q}_{1}\vec{q}_{2})$ determine 
the coupling of the different modes and are prescribed solely in terms of static correlation functions. Within a suitably generalized  convolution approximation to account 
for three-particle correlations the vertex is finally obtained to 
\begin{align}\label{eq:vertex}
& \mathcal{Y}^\alpha_{\mu,\mu_{1} \mu_{2}}(\vec{q},\vec{q}_{1}\vec{q}_{2}) = \nonumber \\
&= \frac{n_0^2}{L^4} \sum_\kappa v_{\mu-\kappa}^* \left[ b^\alpha(
\vec{q}_1 \cdot \vec{q}/q, Q_{\kappa-\mu_2}) c_{\kappa-\mu_2,\mu_1}(q_1) + (1\leftrightarrow 2) \right]. 
\end{align}
The equations of motion, Eqs.~(\ref{eq:eom1},\ref{eq:split},\ref{eq:eom_Newton}), together with the MCT closure, Eq.~\eqref{eq:MCT_ansatz},
 constitute a complete set of equations with unique solutions~\cite{Lang2013b}  if  the static structure is taken as input. 

Here, we focus on  the long-time properties of the intermediate scattering functions
 with particular emphasis on the wavevector-dependent behavior of the long-time limits
\begin{equation}
 F_{\mu\nu}(q) := \lim_{t\to\infty} S_{\mu\nu}(q,t),
\end{equation}
also known as nonergodicity parameters. For the case of bulk liquids it has been proven that the limit exists~\cite{Franosch2014}, and here we assume that this holds also for the case of confined fluids. The nonergodicity parameters are directly measurable quantities  in simulations or experiments and their wavenumber dependence encodes valuable information on the structure of the arrested fluid. 

Glassy states are characterized within the MCT by  non-vanishing nonergodicity parameters $F_{\mu\nu}(q) \neq 0$ while in the liquid state all correlation functions decay to zero. However, in simulations the structural arrest is only transient, 
the coherent intermediate scattering functions for the supercooled regime eventually decay to zero at very long times. In that case the 
frozen-in parts describe the plateau values of the intermediate scattering functions. 

A remarkable property of the theory is that the matrix-valued nonergodicity parameters $F_{\mu\nu}(q)$ can be determined without solving for the full time dependence. Rather the long-time limit of the intermediate scattering function is connected to the corresponding long-time limit of the force kernel as provided by the MCT functional
\begin{equation}\label{eq:static_MCT}
 \boldsymbol{\mathcal{N}}(q) := \lim_{t\to \infty} \boldsymbol{\mathcal{M}}(q,t) = \boldsymbol{\mathcal{F}}[\boldsymbol{F},q].
\end{equation}
A contraction yields a reduced quantity
\begin{equation}\label{eq:static_contraction}
 [\mathbf{N}(q)^{-1}]_{\mu\nu} = \sum_{\alpha,\beta=\parallel,\perp} b^\alpha(q,Q_\mu) [\boldsymbol{\mathcal{N}}^{-1}(q) ]^{\alpha\beta}_{\mu\nu} b^\beta(q,Q_\nu) .
\end{equation}
which considered as functional  of the long-time limits $\mathbf{F}$  displays the  properties of an effective mode-coupling functional~\cite{Lang2012} in the space of matrices with mode-indices $\mu,\nu$. Specializing the equations of motion, Eqs.~(\ref{eq:eom1},\ref{eq:split},\ref{eq:eom_Newton}), to long times, yields the additional relation~\cite{Lang2010,Lang2012}
\begin{equation}\label{eq:fixed_point}
 \mathbf{F} = \mathbf{S} - [\mathbf{S}^{-1} + \mathbf{N}[\mathbf{F}] ]^{-1},
\end{equation}
which has been cast into a form reminiscent of the MCT equations for mixtures~\cite{Franosch2002}.

In general, the fixed-point equations, Eqs. (\ref{eq:static_MCT}),(\ref{eq:static_contraction}), and (\ref{eq:fixed_point}), display many solutions, in particular a vanishing nonergodicity parameter $\mathbf{F}(q)=0$ always constitutes a trivial solution. Since the nonergodicity parameters are long-time limits of correlation functions they have to correspond to positive-semidefinite matrices. One can show~\cite{Lang2012,Lang2013b} that the solution corresponding to the long-time limit of the intermediate scattering functions is maximal and can be obtained within a monotonic iteration scheme. 

The MCT for confined liquids has been extended to include also the tagged-particle motion~\cite{Lang2014a}. The fluctuating density modes of the tagged particle density are given here as
\begin{equation}
 \rho_\mu^{(s)}(\vec{q},t) = e^{i \vec{q} \cdot \vec{r}_s(t)} \exp[ i Q_\mu z_s(t) ] ,
\end{equation}
where $\vec{x}_s = (\vec{r}_s, z_s)$ denotes the position of the tagged particle. The associated
generalized incoherent scattering functions are defined by
\begin{equation}\label{eq:self_intermediate}
 S_{\mu\nu}^{(s)}(q,t) = \langle \rho_\mu^{(s)}(\vec{q},t)^* \rho_\nu^{(s)}(\vec{q}) \rangle. 
\end{equation}
The corresponding equations of motion are in essence identical to the one of the collective motion, and will not be repeated here. The mode-coupling ansatz represents the associated force kernel in terms of products of the collective and the incoherent scattering functions with vertices that encode again only structural properties, for explicit expressions see Ref.~\cite{Lang2014a}. Hence, in order to calculate the tagged-particle correlators, the equations for the coherent motion need to be solved as input.

\section{Simulations and numerical results}
\label{sec:results}

Here we describe the set-up of the computer simulation for a polydisperse hard-sphere fluid in confinement, next we compare the simulations to numerical results of the fundamental-measure theory as well as the Percus-Yevick theory for the generalized static structure factors. We calculate the
nonergodicity parameters within mode-coupling theory and provide a qualitative comparison to the simulational results both for the coherent as well as for the self dynamics.

\begin{figure}[htp]
(a) \includegraphics*[width=0.9\linewidth]{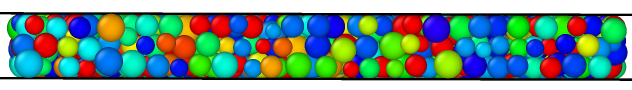} \\
(b) \includegraphics*[width=\linewidth]{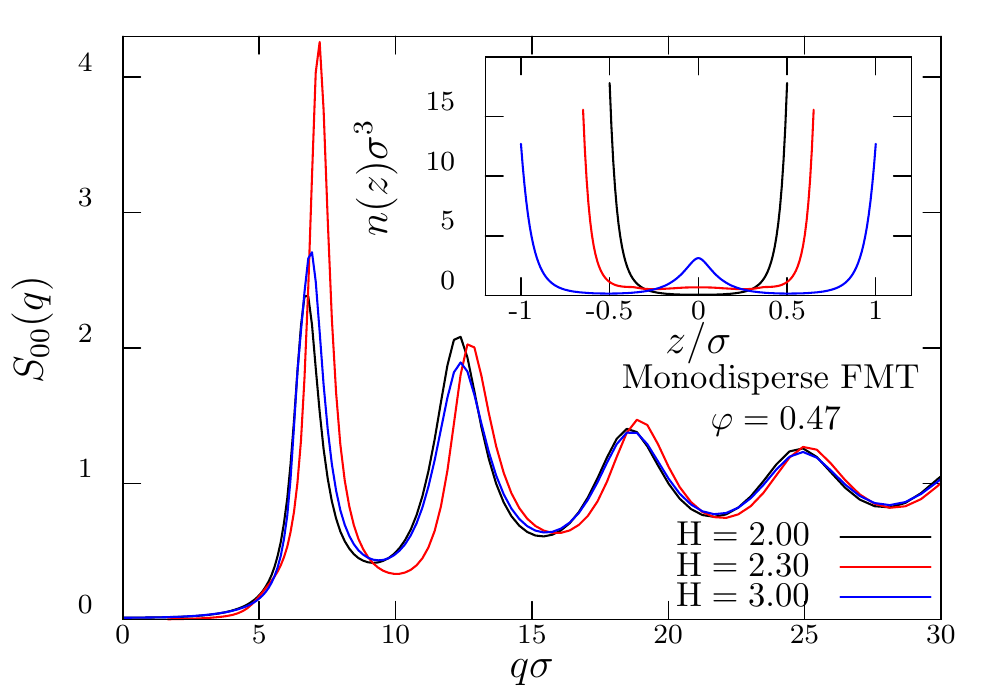}\\
(c) \includegraphics*[width=\linewidth]{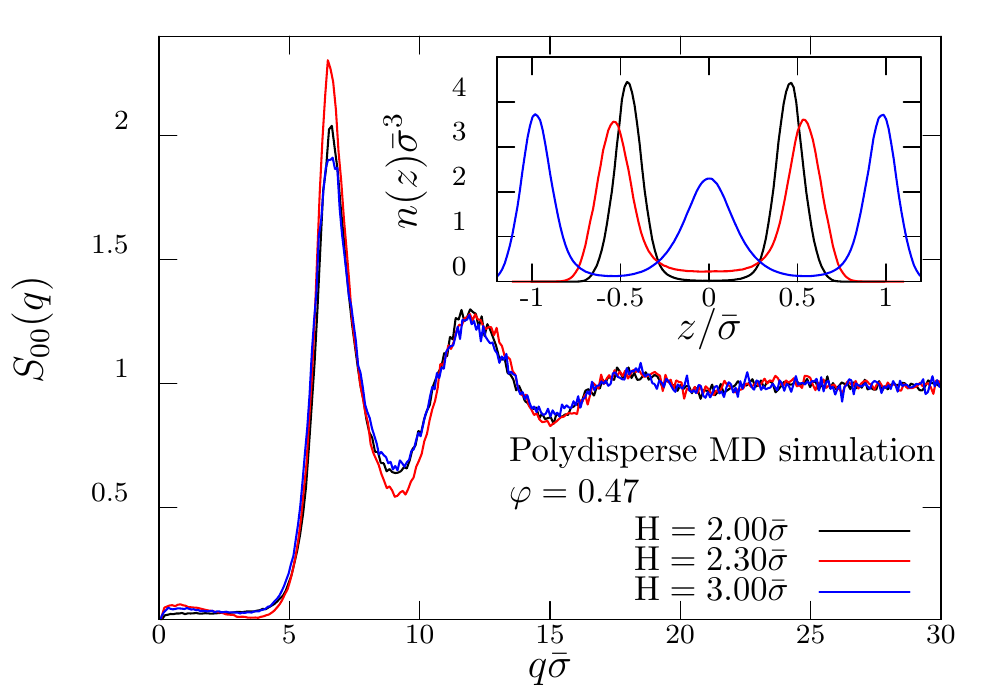}
\caption{(a) Snapshot of a confined polydisperse hard-sphere liquids for $H=3.0 \bar \sigma$.
(b) Structure factor $S_{00}(q)$ obtained from the inhomogeneous Percus-Yevick theory at fixed $\varphi=0.47$. Inset: density profiles $n(z)$ for different slit widths as obtained from FMT. (c) Simulated static structure factor $S_{00}(q)$ for different plate distances $H$ at packing fraction $\varphi=0.47$. Inset: simulated density profiles for various wall-to-wall distances at the same packing fraction. The first sharp diffraction peak varies non-monotonically; lowest for $H=2.0\bar\sigma$ and $H=3.0\bar\sigma$, highest for $H=2.3\bar\sigma$.}
\label{fig:sq00_plus_density}
\end{figure}

\subsection{Simulations}
\label{sec:simulation}
We have performed extensive event-driven molecular dynamics (EDMD) simulations for  hard-sphere systems. To mimic the experimental set-up and to circumvent 
crystallization we use a slightly polydisperse system. The particle size distribution has been drawn from a Gaussian around a mean diameter of $\bar \sigma$ with a polydispersity (standard deviation)  of 15\%. The particles are confined between two planar hard walls placed in parallel at
physical distances  $\pm H/2$ from the slit center and periodic boundary conditions are imposed along the lateral directions. 
The centers of the particles cannot come closer to the walls than their respective hard-sphere exclusion radius, i.e. on average the accessible slit width corresponds to $L = H-\bar{\sigma}$.   

A  snapshot of a dense hard-sphere system in  slit geometry  is shown in Fig.~\ref{fig:sq00_plus_density}a. The volume of the simulation box is $V=L_{\text{box}}^2 H$, where the lateral system size $L_{\text{box}}$ varies in the range from $60\bar\sigma$ to $75\bar\sigma$. Depending on $H, L_{\text{box}}$ and the packing fraction $\varphi = (N/V) \pi \overline{\sigma}^3/6 $, the number of particles $N$ ranges between $8\, 000$ and $30\, 000$.  
Due to the hard-sphere interaction, the thermal energy  $k_B T$ enters only via the time scale  $t_{0}=\sqrt{m\bar\sigma^{2}/k_{B}T}$. 
These confined liquids have been equilibrated via
 long simulations, extending up to 5 decades in time, such that particles have traversed distances larger than the microscopic 
cage length and the corresponding mean-squared displacements have reached the diffusive regime. Furthermore we have checked that no ageing occurs. Production runs have been performed from equilibrated configurations only to ensure that all data correspond to equilibrium dynamics. 

To measure the coherent as well as the self intermediate scattering  functions 200 independent runs 
for $H=2.0\bar \sigma$, $H=2.3\bar{\sigma}$, and $H=3.0\bar\sigma$ at $\varphi=0.52$ have been performed. We also checked that the results are free from segregation or finite size effects.

\subsection{Static properties}
\label{subsec:static}

The qualitative agreement between theory and simulation at the static level is essential to test the MCT predictions via computer simulations. Here we calculate the equilibrium density profile in the slit by minimizing explicitly the grand potential functional within fundamental-measure theory (FMT) with the White Bear version II for the excess free energy functional~\cite{Roth2010,Hansen2006}. For monodisperse hard-sphere system, the minimization condition leads to the following equation

\begin{equation}
  \ln n(z)=\beta \mu -\beta \frac{\delta F^{\text{ex}}[n]}{\delta n(z)} - \beta V(z),
\end{equation}
where $n(z)$ is the particle number density with diameter $\sigma$, $\beta= (k_{B}T)^{-1}$ is the inverse thermal energy, $\mu$ is the chemical potential specified by the particle reservoir, $V(z)$ is the external potential, and $F^{\text{ex}}[n]$ is the excess free-energy functional from FMT~\cite{Roth2010,Hansen2006}. It has been demonstrated that FMT gives very precise density profiles for high densities of the hard-sphere fluid in various geometries. A stable numerical solution was obtained iteratively in Fourier space with a grid resolution up to $0.001\sigma$ and is shown in the inset of Fig.1(b).

Fundamental-measure theory predicts an oscillatory density profile, $n(z)$, for monodisperse systems along the direction perpendicular to the walls, see the inset of Fig.~\ref{fig:sq00_plus_density}(b). The FMT clearly suggests an accumulation of particles close to the walls, $z=\pm H/2$. The corresponding simulational $n(z)$, see the inset of Fig.~\ref{fig:sq00_plus_density}(c), shares also the same oscillatory density profile, although the peaks near to the walls are less pronounced. In particular the peak is not cut off at closest distance of the average particle, since in a polydisperse system smaller particles can come closer to the walls than larger ones. 
The differences to the theory are solely due to polydispersity, a full agreement  can be achieved  by evaluating FMT for  mixtures of hard-sphere particles of up to 51 particle radii to mimic a polydisperse system~\cite{Mandal2014b}. Note, that in the MCT calculations the density profile enters in terms of its Fourier coefficients, Eq.~\eqref{eq:density}, the lowest-order modes being the most important. Therefore small 
details of the density profile should not make a qualitative difference for the comparison of MCT results to simulations.

The static structure factors in confined geometry encode two-particle correlations which can be calculated in liquid-state theory by suitable closures of the inhomogeneous Ornstein-Zernike relation~\cite{Hansen:Theory_of_Simple_Liquids,Henderson:Fundamentals_of_inhomogeneous_fluids,Ram2014}. For hard spheres the Percus-Yevick theory (PY) has been shown to yield a successful description also in slit geometry~\cite{Nygard2012,Nygard2013, Lang2010, Mandal2014b}. We have solved numerically the PY equations relying on our decomposition into Fourier modes, Eq.~\eqref{eq:OZ_equation}, using the density profile obtained from FMT as input. In principle the direct correlation function determines  the density profile via the Lovett-Mou-Buff-Wertheim equation~\cite{Nygard2013,Henderson:Fundamentals_of_inhomogeneous_fluids}. Our combined FMT-PY results violate this exact relation, however since the FMT provides reliable results for the density profile the differences are expected to be small. 
 In Fig.~\ref{fig:sq00_plus_density}(b) we present the slit width dependence of the lowest mode 
$S_{00}(q)$ at fixed packing fraction $\varphi=0.47$. Since 
 $S_{00}(q)$ includes only modulations parallel to the walls we refer to it as the in-plane static structure factor. The overall 
shape of $S_{00}(q)$ for different slit widths is similar to bulk liquids, and the oscillations persist all the way to large wavenumbers. 
 These results reveal a  nonmonotonic steep shoot-up of the first sharp diffraction peak as the plate distance is varied. 
For the distances investigated, the maximum appears for $H \approx 2.3\bar{\sigma}$, i.e. at incommensurate packing of the particles in the slit. 
The enhancement of the first sharp diffraction peak reflects that the layers in the slit are strongly coupled and particles cannot pass each other due to steric constraints. For bulk liquids the primary peak of the structure factor has been identified as pivotal to induce the structural arrest, although 
other features such as the behavior at large wavenumbers can be crucial for higher-order glass transitions~\cite{Gotze_Complex_Dynamics}. Since the lowest-order structure displays nonmonotonic behavior we anticipate that the nonequilibrium state diagram also displays an oscillatory glass-transition line. In simulations, we calculate $S_{00}(q)=S_{00}(q,t=0)$ as described in Eqn.~\eqref{eq:scattering_function} using the particle positions. The maximum of the peak also occurs at the same wall separation in simulations, however, the peaks are  less pronounced and the oscillations die away after the second peak due to polydispersity. Both theory and simulations exhibit the progressive structuring for incommensurate packing and similarly destructuring for commensurate packing along lines of  constant packing fraction.

\begin{figure}[htp]
(a)\includegraphics*[width=\linewidth]{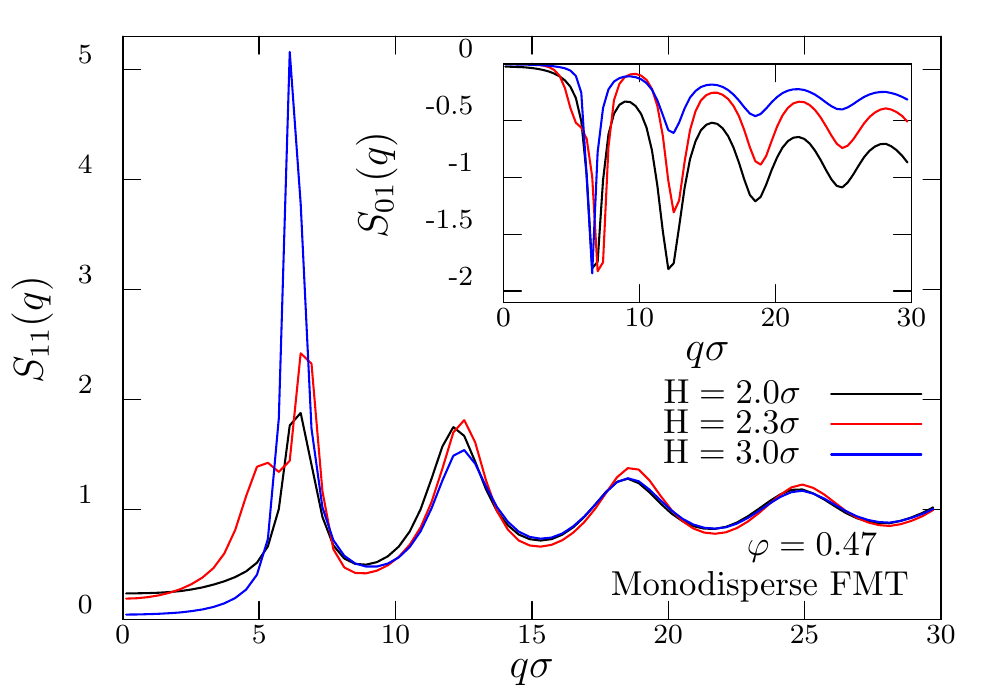}
\\
(b)\includegraphics*[width=\linewidth]{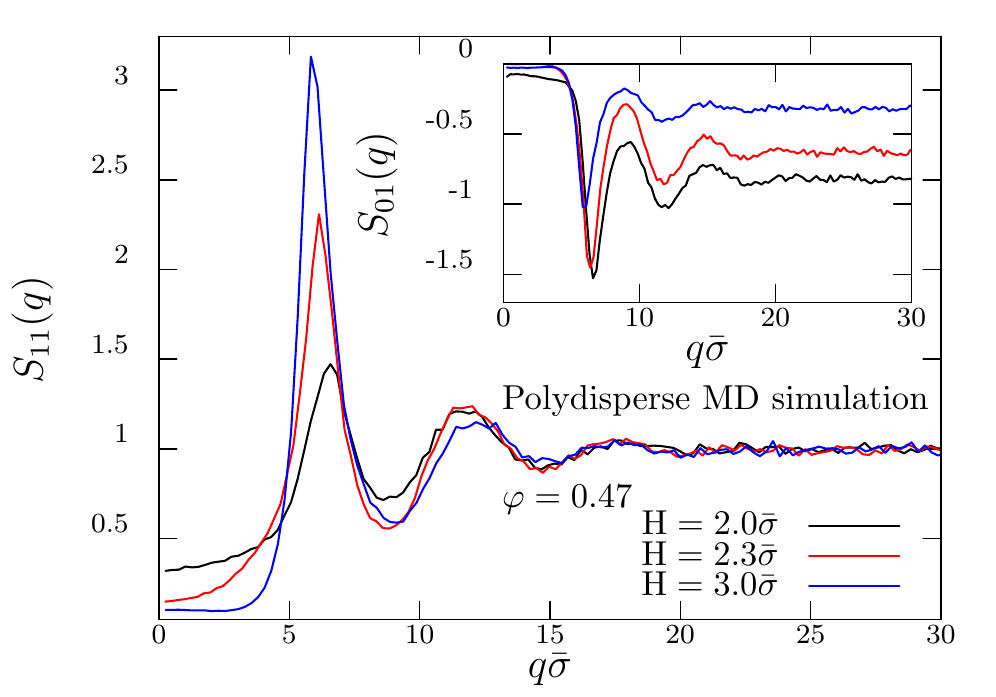}
\caption{(a) The static structure factor $S_{11}(q)$ using the inhomogeneous PY closure for different plate distances $H$ at packing fraction $\varphi=0.47$. Here, the first sharp diffraction peak varies monotonically as a function of the slit width. Inset: non-diagonal structure factor $S_{01}$ obtained from FMT. (b) Simulated structure factor $S_{11}(q)$ for $\varphi=0.47$. Inset: simulated $S_{01}(q)$ at the same packing fraction.}
\label{fig:sq11_plus_01}
\end{figure}

In Fig.~\ref{fig:sq11_plus_01} we present the slit width dependence of $S_{11}(q)$, which is the first component of the generalized static structure factor sensitive to the arrangement of particles in the direction perpendicular to the walls. One infers that the first diffraction peak of $S_{11}(q)$ exhibits a very different slit-width dependence in both simulation and theory compared to the in-plane static structure factor $S_{00}(q)$.
Here, the first sharp  diffraction peak appears to grow monotonically as the slit width is increased. For very large slit widths the generalized static structure factor is expected to become diagonal~\cite{Lang2014b} 
\begin{equation}
 S_{\mu\nu}(q) \to \delta_{\mu\nu} S(k) , \qquad \text{with} \quad k^2 = q^2 + Q_\mu^2, 
\end{equation}
and indeed, already for $H= 3.0\sigma$ the calculated $S_{11}(q)$ resembles the in-plane structure factor. The first nontrivial off-diagonal static structure factor $S_{01}(q)$  also displays characteristic oscillations anticorrelated to the ones of the in-plane static structure factor $S_{00}(q)$, see 
 inset of Fig.~\ref{fig:sq11_plus_01}. These oscillations fade out as the plate separation increases.


\subsection{Numerical implementation of  the MCT fixed-point equations}
\label{subsec:mct_numerics}

\begin{figure}[htp]
(a)\includegraphics*[width=\linewidth]{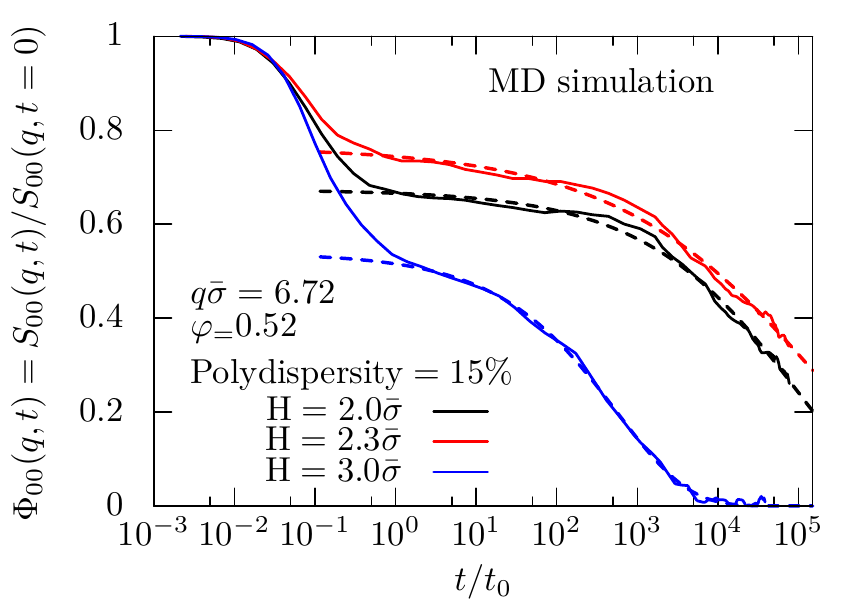}\\
(b)\includegraphics*[width=\linewidth]{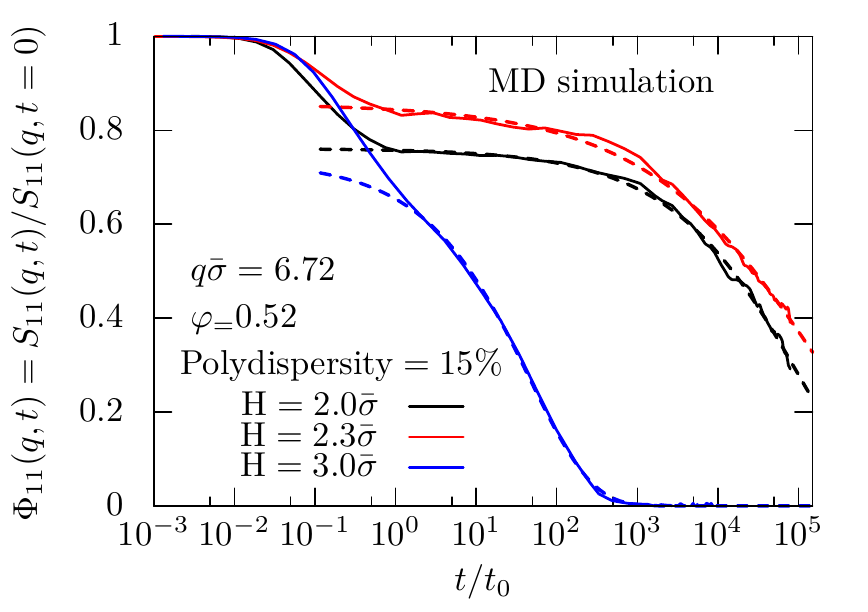}
\caption{Normalized collective intermediate scattering functions $\Phi_{00}(q,t)$ in (a) and $\Phi_{11}(q,t)$ in (b) for different values of $H$ at fixed $\varphi=0.52$ and wavenumber $q \bar{\sigma} = 6.72$. Both coherent scattering functions exhibit a non-monotonic relaxation as a function of the slit width.The dashed lines are KKW fits, Eq.~\eqref{eq:kwwc}.}
\label{fig:csf}
\end{figure}

To find the nonergodicity parameter for the distances $H=2.0\sigma, 2.3\sigma,\text{and}   $ $3.0\sigma$, the Eqs.~\eqref{eq:static_MCT},\eqref{eq:static_contraction}, and \eqref{eq:fixed_point} have been solved by iteration. The discrete mode-indices have been 
truncated to  $|\nu| \leq 10$ and the wavevectors have been discretized on a grid  $q=\hat{q}\Delta q + q_{0}$ with parameters $q_{0}\sigma=0.1212,{\Delta q}\sigma=0.4$ and grid range $\hat{q}=0,1,\dots N-1$ with $N=150$. For the  plate separations chosen the perpendicular wavenumbers extend to at least $ Q_\nu \sigma \simeq 30$. 
To reduce the computing time and complexity we have retained only the diagonal elements of matrix-valued quantities. For example, in  Eq.~\eqref{eq:vertex} the direct correlation function is replaced by a diagonal matrix $c_{\mu\nu}(q) \mapsto c_{\mu\mu}(q) \delta_{\mu\nu}$, and similarly we keep only $v_0^* \neq 0$ for the vertex. This entails that only couplings $\mu = \mu_1 + \mu_2$ are included in    
 the MCT functional. Furthermore  
$\mathcal{N}^{\alpha \beta}_{\mu\nu}(q)$, Eq.~\eqref{eq:static_MCT}
 is treated as non-vanishing only for $\alpha=\beta$ and $\mu=\nu$, which implies that the contracted quantities $N_{\mu\nu}(q)$, 
Eq.~\eqref{eq:static_contraction}
 are also diagonal in the mode indices $\mu,\nu$. 
Similarly,  in Eq.~\eqref{eq:fixed_point} the static structure factor $S_{\mu\nu}$ 
is replaced by a diagonal matrix.  As a consequence  the coupling of the nonergodicity parameters arises solely on the level of the mode-coupling functional, i.e. ${\cal N}^{\alpha\alpha}_{\mu\mu}(q)$ requires $F_{\mu_1\mu_1}(q_1)$ as input for all modes $\mu_1$ and wavenumbers $q_1$. Since the couplings involve the direct correlation functions, one anticipates that the mode $c_{00}(q)$ plays a dominant role for the MCT solutions.

We have utilized fundamental-measure theory~\cite{Hansen2006,Roth2010} to evaluate $n(z)$ and have employed a Percus-Yevick approximation to close the inhomogeneous Ornstein-Zernike relation~\cite{Henderson:Fundamentals_of_inhomogeneous_fluids,Nygard2012, Nygard2013} to solve for the structure factors. Using the matrix-valued static structure factors for different slit width as input to Eqs.~\eqref{eq:static_MCT},\eqref{eq:static_contraction}, and \eqref{eq:fixed_point}, it is found that either $\mathbf F(q)$ is zero for all $q$ or $\mathbf F(q)$ is nonzero and does not change any longer for all $q$. We  compare  the  nonergodicity parameters for $H=2.0\sigma$, $H=2.3\sigma$, and $H=3.0\sigma$ in the next subsection~\ref{subsec:coherent_nep}. For this purpose, we select a packing fraction $\varphi=0.47$ such that the MCT equations yield glassy states for all slit widths.
 
\subsection{Coherent nonergodicity parameters from MCT and simulations}
\label{subsec:coherent_nep}


In this subsection we exemplify the nonergodicity parameters for the collective motion for our confined hard-sphere system. It is known from bulk systems that  
the MCT underestimates the critical packing fraction by typically 20\%, hence to observe the signature of the glass transition in simulations, we have increased
 the packing fraction to $\varphi=0.52$. 
The initial values $S_{\mu\nu}(q,t=0)$ correspond to the static structure factors that have been discussed in Subsec.~\ref{sec:simulation}. Here we focus on  the evolution of the structural relaxation with varying slit width and present normalized 
scattering functions $\Phi_{\mu\nu}(q,t)=S_{\mu\nu}(q,t)/S_{\mu\nu}(q,0)$ at constant packing 
fraction.  Furthermore we restrict the discussion to the lowest  diagonal components of the intermediate scattering functions. 
In Fig.~\ref{fig:csf} we display $\Phi_{00}(q,t)$ and $\Phi_{11}(q,t)$ corresponding to a wavenumber of $q\bar{\sigma} =6.72$ close to the first sharp diffraction peak for different plate separations. After an initial decay which appears to be independent of the slit width, an extended
 plateau is reached at intermediate time scales, followed by a pronounced non-exponential relaxation. The plateau value depends in a nonmonotonic fashion on the wall distance. Upon increasing the slit width from the closest commensurate distance $H = 2.0 \bar{\sigma}$ to the incommensurate value $H= 2.3\bar{\sigma}$ the plateau increases, signalling a stronger frozen-in structure, while the structural relaxation times remain close to each other.  Widening the slit further  to the commensurate value $H= 3.0 \bar{\sigma}$  the structural relaxation speeds up  by two orders of magnitude, concomitantly the plateau value decreases significantly in $\Phi_{00}(q,t)$.


To systematically extract the plateau values corresponding to the nonergodicity parameters of the theory, we rely on fits to a phenomenological Kohlrauch-William-Watts (KWW) stretched exponential~\cite{Williams1970} 
\begin{equation}
 \Phi_{\mu\nu}(q,t)=f_{\mu\nu}(q)\exp(-(t/{\tau})^{\beta^{c}}),
\label{eq:kwwc}
\end{equation}
where $\beta^{c}$ is the Kohlrausch exponent, $\tau$  the relaxation time, and $f_{\mu\nu}$ is our estimate for the  coherent nonergodicity parameters.
These values depend slightly on the time windows chosen, in particular, for states where the glassy relaxation is poorly developed. In order to have a consistent set, we have used the same time window $1.0 \lesssim  t/t_0 \lesssim10^5$ and employed  a least-square fitting routine. 

\begin{figure}[htp]

\includegraphics*[width=\linewidth] {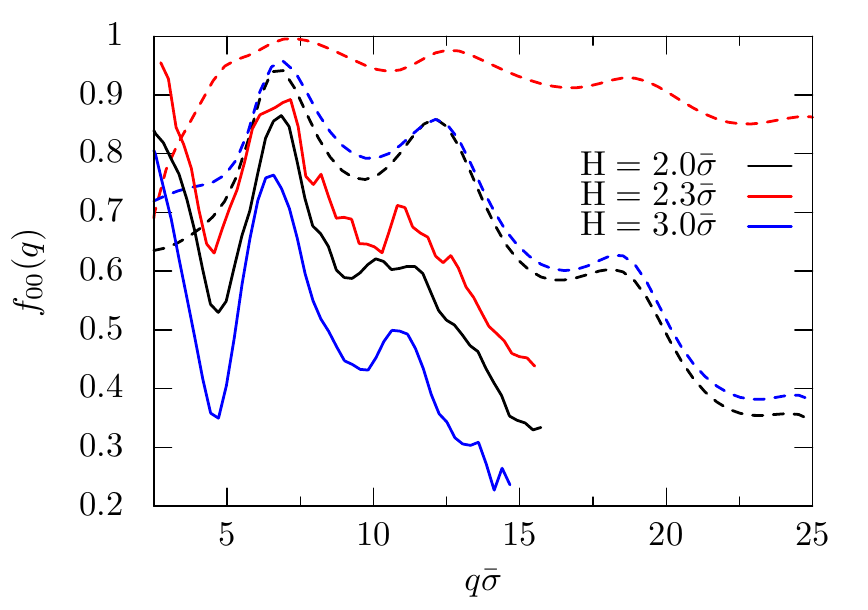}
\caption{Normalized wave-vector dependent coherent nonergodicity parameters $f_{00}(q)$ by solving the MCT Eq.~\eqref{eq:fixed_point} for different values of $H$ at $\varphi=0.47$ (dashed lines). Simulated $f_{00}(q)$ by fitting Eq.~\eqref{eq:kwwc} for three 
different values of $H$ at $\varphi=0.52$ (solid lines). Since the slow dynamics of polydisperse computer simulations 
at $\varphi=0.47$ is not significant, we increase the packing fraction to $\varphi=0.52$. }
\label{fig:nep_coherent_f00}
\end{figure}

The coherent  nonergodicity parameters $f_{00}(q)$ evaluated from MCT solutions are displayed in Fig.~\ref{fig:nep_coherent_f00}(a) as a function of wavenumber and compared to the ones extracted via the KWW fits in Fig.~\ref{fig:nep_coherent_f00}(b). The variation with slit width reflects the behavior of the corresponding static structure factors. The dependence on wavenumber is similar to the variations of the associated structure factors, in particular, the nonergodicity parameters display oscillations in phase with the structure factors. Therefore the rule of thumb valid for bulk systems that the oscillations of the structure factor are reflected also in the \emph{normalized} nonergodicity parameters appears to hold also in confinement for the lowest mode $f_{00}(q)$. The nonmonotonic behavior as the slit width is gradually increased is apparent for all wavenumbers. The nonergodicity parameters obtained from MCT
for the incommensurate case suggest a much stronger structural arrest than the ones extracted from the simulation, which could be due to the polydispersity. Second, the long-wavelength limit of the nonergodicity parameters in simulations is much higher than the MCT prediction. Third, again due to the polydispersity in our simulations the nonergodicity parameters for $H=3.0\bar \sigma$ always lie below the nonergodicity parameters for $H=2.0\bar \sigma$, see the corresponding variation of the in-plane static structure factor in Fig.~\ref{fig:sq00_plus_density}(c).
A similar observation has been made for colloidal bulk liquids and rationalized by polydispersity~\cite{Weysser2010}, in essence the long-wavelength behavior acquires an admixture from the incoherent dynamics. Wavenumbers higher than $q\bar{\sigma} \approx 15$  have been omitted from the figure since the simulational data become noisy and the KWW fits are unreliable.

%

\begin{figure}[htp]
\centering
\includegraphics*[width=\linewidth] {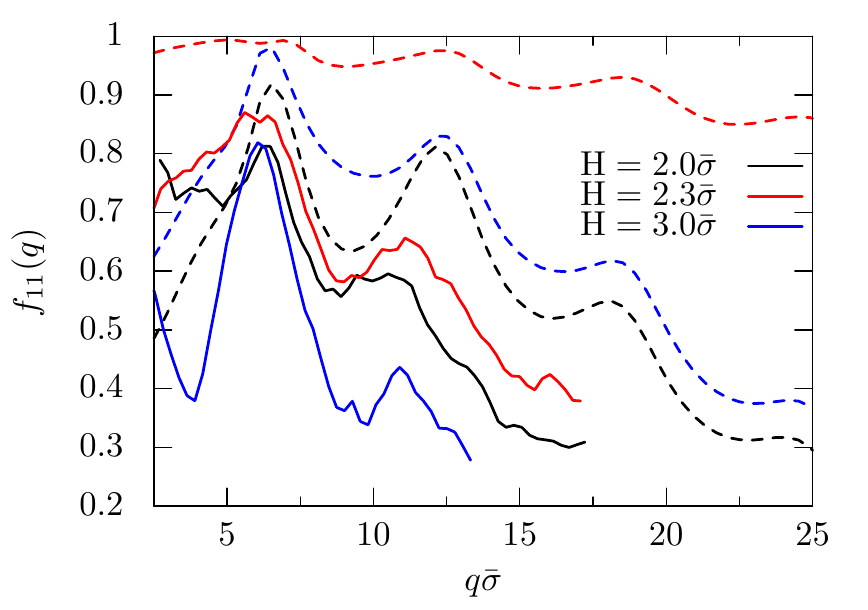}
\caption{Normalized wave-vector dependent coherent nonergodicity parameters $f_{11}(q)$ by solving the MCT Eq.~\eqref{eq:fixed_point} for different values of $H$ at $\varphi=0.47$ (dashed lines). Simulated $f_{11}(q)$ by fitting Eq.~\eqref{eq:kwwc} for three different values of $H$ at $\varphi=0.52$ (solid lines).}
\label{fig:nep_coherent_f11}
\end{figure}

The wavevector dependence of the first higher mode  $f_{11}(q)$ for both MCT calculation as well as the simulation is displayed in Fig.~\ref{fig:nep_coherent_f11} for the three slit widths considered. Both exhibit again a nonmonotonic behavior, similar to the in-plane structure factor $S_{00}(q)$, but in contrast to the initial value of the correlation functions $S_{11}(q)$. In fact, the wavenumber dependence of $f_{11}(q)$ is qualitatively very similar to the in-plane mode $f_{00}(q)$, the oscillations occur at the same wavenumbers as the in-plane structure. 
These results suggest that the 
in-plane structure factor $S_{00}(q)$ is the relevant one for the  particle dynamics, in particular the first sharp diffraction peak appears to be the key ingredient. The differences between MCT prediction and simulational results are restricted to  the long-wavelength behavior.  Interestingly, the value for the long-wavelength limit of the 
nonergodicity parameter for the incommensurate plate separation is higher than the MCT prediction for the in-plane mode $f_{00}(q\to 0)$, while for $f_{11}(q\to 0)$ we observe the opposite behavior.   

We also have inspected the nonergodicity parameters without normalizing the correlation functions, yet then no clear trends can be extracted. The use of  normalized nonergodicity parameters has been useful also for bulk mixtures to separate the changes from the initial value and from the tendency to structural arrest~\cite{Voigtmann2003p}.

\subsection{Nonergodicity parameters of the self motion}
\label{subsec:incoherent_nep}


\begin{figure}[h]
(a)\includegraphics*[width=\linewidth]{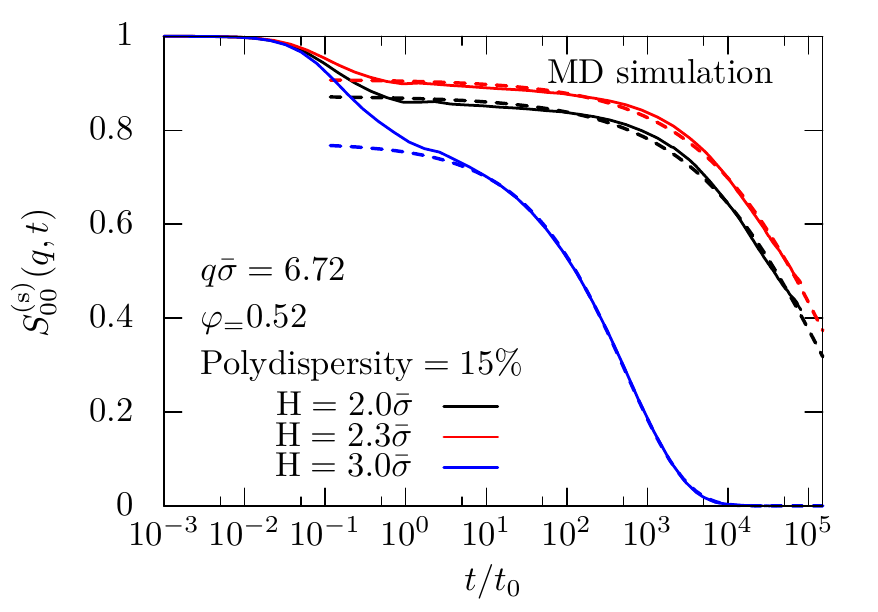}\\
(b)\includegraphics*[width=\linewidth]{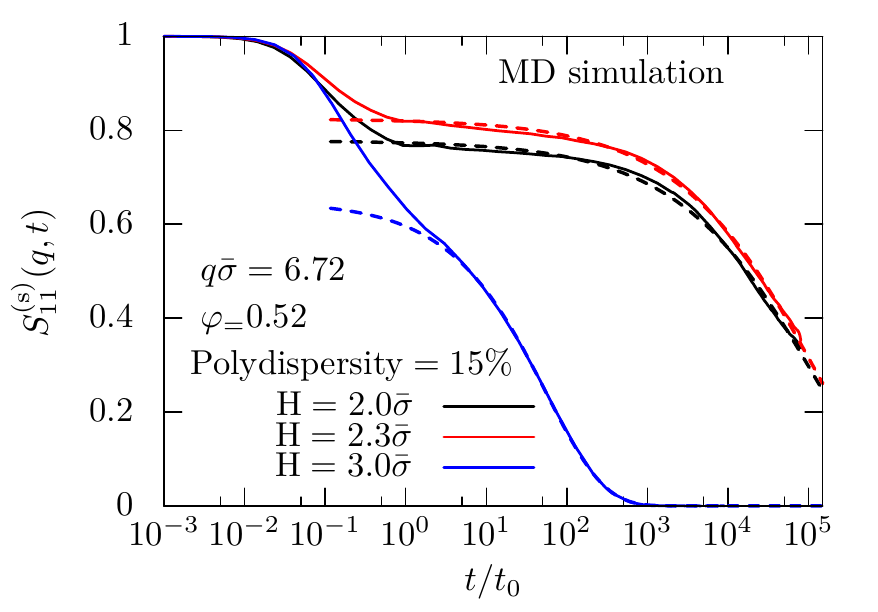}
\caption{Incoherent intermediate scattering functions (a) $S_{00}^{(s)}(q,t)$  and (b) $S_{11}^{(s)}(q,t)$  for different values of the plate separation $H$ at fixed packing fraction $\varphi=0.52$. Both incoherent scattering functions again exhibit a nonmonotonic relaxation as a function of the slit width. The dashed lines are KKW fits, Eq.~\eqref{eq:kwwc}.
}
\label{fig:isf}
\end{figure}

\begin{figure}[h]

\includegraphics*[width=\linewidth]{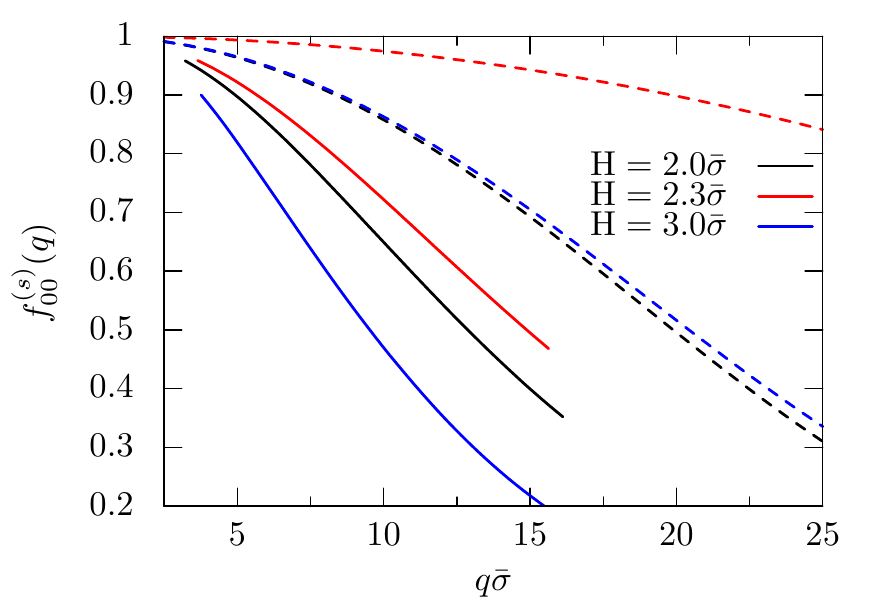}
\caption{Wavenumber dependence of the incoherent nonergodicity parameters $f^{(s)}_{00}(q)$ obtained from MCT for different values of $H$ at $\varphi=0.47$ (dashed lines). Corresponding  $f^{(s)}_{00}(q)$ extracted from simulations  for three different values of $H$ at $\varphi=0.52$ (solid lines).}
\label{fig:self_nep_00}
\end{figure}

\begin{figure}[h]
\includegraphics*[width=\linewidth]{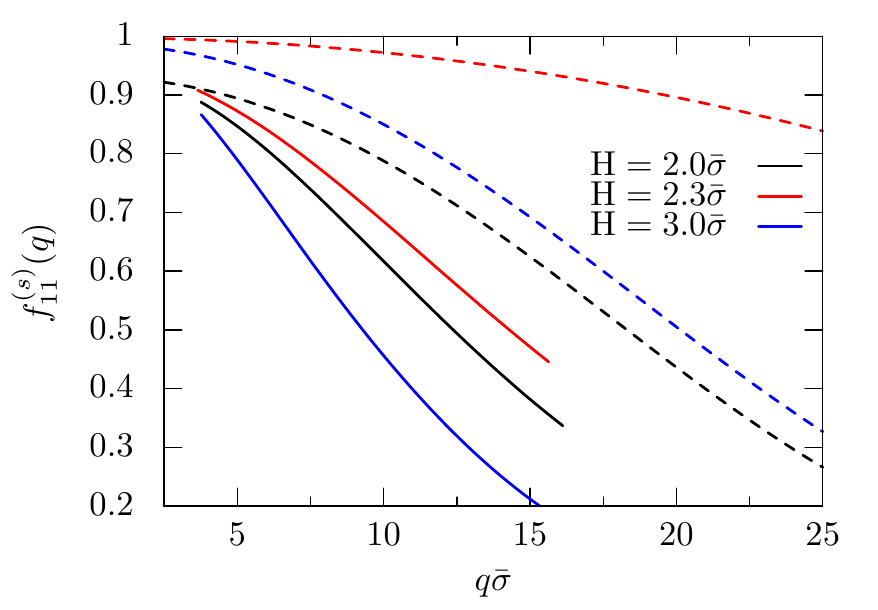}
\caption{Wavenumber dependence of the incoherent MCT nonergodicity parameters $f^{(s)}_{11}(q)$ for different values of $H$ at $\varphi=0.47$ (dashed lines). Corresponding measured  $f^{(s)}_{11}(q)$ for the three different values of $H$ at $\varphi=0.52$ (solid lines).}
\label{fig:self_nep_11}
\end{figure}

\begin{figure}[h]
\includegraphics*[width=\linewidth]{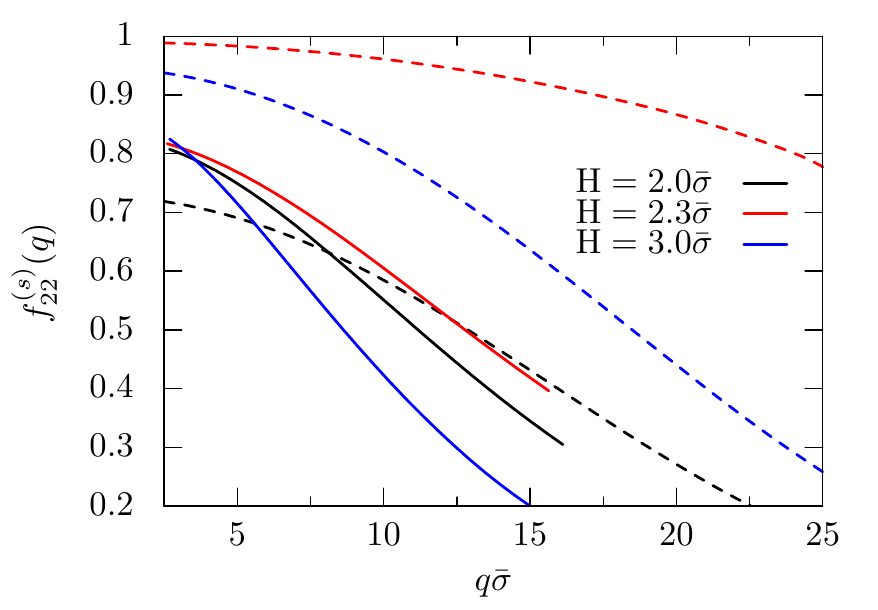}
\caption{Wavenumber dependence of the incoherent MCT nonergodicity parameters $f^{(s)}_{22}(q)$ for different values of $H$ at $\varphi=0.47$ (dashed lines). Corresponding measured  $f^{(s)}_{22}(q)$  for the three different values of $H$ at $\varphi=0.52$ (solid lines).}
\label{fig:self_nep_22}
\end{figure}

%
%
%
%
%

The simpler quantity from a simulational point of view is the self motion or the incoherent intermediate scattering functions as prescribed in Eq.~\eqref{eq:self_intermediate}, since the statistics is significantly enhanced by averaging over all particles. In contrast, from the perspective of the MCT the self motion is a  derived quantity, the equations for the collective need be evaluated first. The MCT numerics relies on the same truncation of mode indices and the diagonal approximation.

The shape of the measured  incoherent intermediate scattering functions $S^{(s)}_{00}(q,t)$ and $S^{(s)}_{11}(q,t)$ for different values of $H$, see ~Fig.~\ref{fig:isf}, are in close resemblance to the coherent one, yet with a significantly smaller statistical error. In particular, the self motion reflects the nonmonotonic behavior of the dynamics. A normalization is not necessary here, since the initial values $S_{\mu\nu}^{(s)}(q,t=0)$ of the diagonal components are unity anyway~\cite{Lang2014a}.

To extract the plateau values of the incoherent scattering functions we rely again on a KWW fit, similar to Eq.~\eqref{eq:kwwc}. The resulting  incoherent nonergodicity parameters $f^{(s)}_{00}(q)$, $f^{(s)}_{11}(q)$, and $f^{(s)}_{22}(q)$  from simulations are displayed in Figs.~\ref{fig:self_nep_00}, \ref{fig:self_nep_11}, and \ref{fig:self_nep_22} together with the  corresponding MCT predictions. The nonmonotonic  behavior is reflected for all wavenumbers, the curves are all bell-shaped, in contrast to their coherent counterparts. For the lowest mode, they extrapolate to unity for vanishing wavenumber, which merely reflects particle conservation. For the higher modes a value smaller than unity is anticipated characterizing the freezing in the transverse direction. The simulational data for small
 wavenumbers are difficult to obtain due to the finite size of the system, yet the data suggest that the long-wavelength limit indeed differs from unity. Let us mention that the shape of the incoherent diagonal nonergodicity parameters is similar to the one of a tagged molecule in a simple liquid where the mode indices refer to orientational degrees of freedom~\cite{Franosch1997}. In principle, fitting $f^{(s)}_{00}(q)$ to a Gaussian curve $\exp(-q^{2}{r_{c}}^{2})$ provides an estimate of the localization length $r_c$.  It is clear that this localization length is decreased from commensurate packing $H=2.0 \bar \sigma$ to incommensurate packing $H=2.3 \bar \sigma$.

\begin{figure}[htp]
(a)\includegraphics*[width=0.82\linewidth]{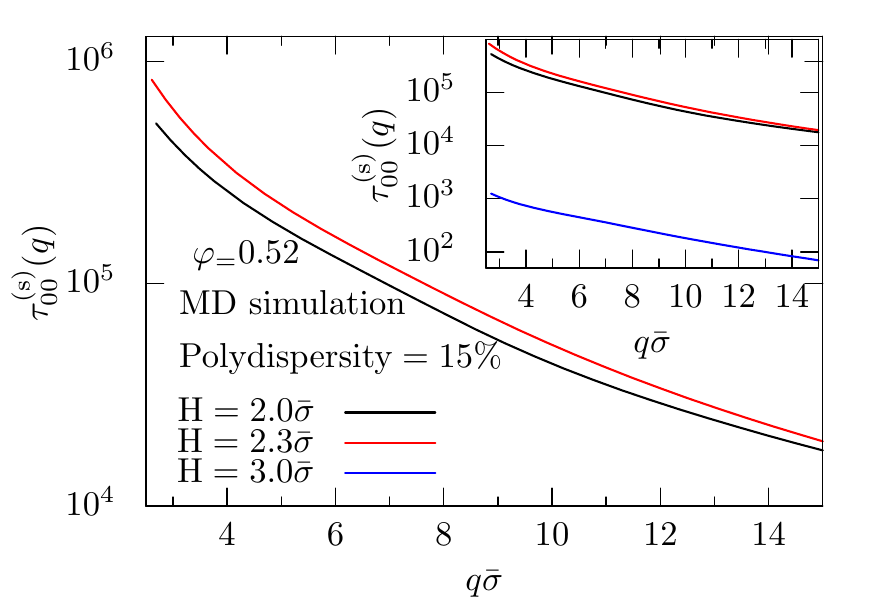}\\
(b)\includegraphics*[width=0.82\linewidth]{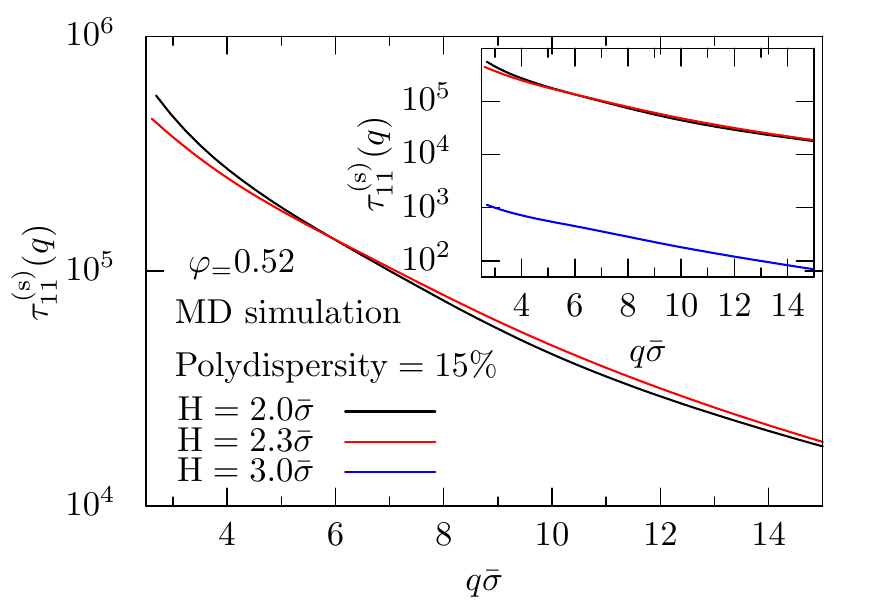}
\caption{(a) Self $\alpha$-relaxation times $\tau_{00}^{(s)}(q)$ as function of the wavenumber $q\bar \sigma$ for $H=2.0\bar \sigma$ and $H=2.3\bar \sigma$ at $\varphi=0.52$. The inset includes an additional wall distance $H=3.0\bar \sigma$. Higher mode $\tau_{11}^{(s)}(q)$ is presented in (b).}
\label{fig:self_relaxation}
\end{figure}

\begin{figure}[htp]
(a)\includegraphics*[width=0.82\linewidth]{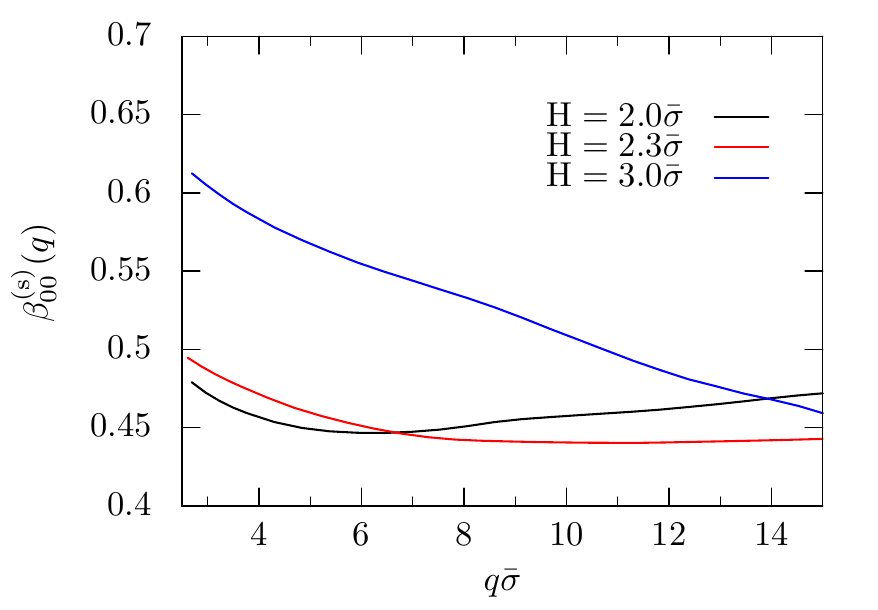} \\
(b)\includegraphics*[width=0.82\linewidth]{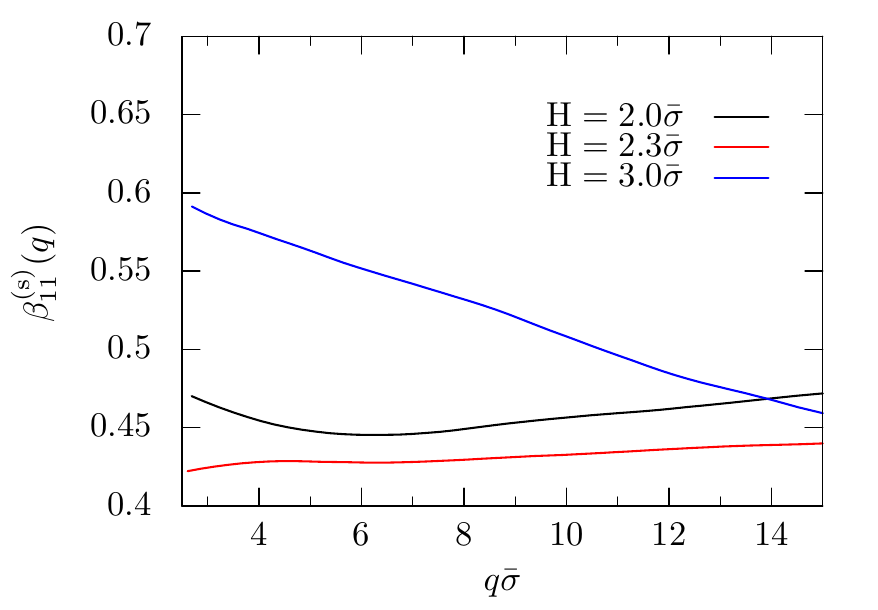}
\caption{Self Kohlrausch stretching exponents $\beta_{00}^{(s)}$ and $\beta_{11}^{(s)}$ as a function of wavenumber $q\bar \sigma$ in (a) and (b) respectively for differenet slit widths, determined from fits to the simulation data using Eq.~\eqref{eq:kwwc}.}
\label{fig:beta_rexponent}
\end{figure}

The wavenumber-dependent relaxation times extracted from the simulations reflect again the non-monotonic dependence on the slit width, see Fig.~\ref{fig:self_relaxation}. Interestingly, the curves for the higher modes display intersections as the plate separation is changed. We also compute the Kohlrausch exponents $\beta_{00}^{s}$ and $\beta_{11}^{s}$ as a function of wavenumber for different slit widths, as shown in Fig.~\ref{fig:beta_rexponent}. The curve for the incommensurate wall separation is comparatively more stretched, which indicates stronger heterogeneous dynamics inside the slit. 

\section{Summary and Conclusion}
\label{sec:conclusion}

In this work  we have reported an extensive study of confined liquids  in terms of theory and simulations. We have focused on the correlations between the structure and dynamics of dense liquids in planar confinement. Therefore we have solved for  the long-time limits of the  MCT equations 
for confined liquids to obtain both the coherent and the incoherent nonergodicity parameters. A second goal  of this study was to test the MCT predictions for the dynamics by means of computer simulations. After the successful test of the matrix-valued structure factors using the inhomogeneous PY closure, we have computed the $q$-dependence of the nonergodicity parameters for different slit widths at a common packing fraction. Strikingly, the MCT predicts a nonmonotonic behavior of the nonergodicity parameters as a function of the slit width at fixed packing fraction. These results have also been complemented by our simulations. Our study suggests that the nonmonotonic behavior of both the coherent and the incoherent 
nonergodicity parameters can be understood as a dynamic manifestation of the switching between commensurate and incommensurate packing upon 
variation of the slit width. Roughly speaking, a commensurate packing allows for a more disordered local structure  of confined liquids, which favors an efficient sliding motion parallel to the walls. In contrast, an incommensurate packing obstructs the sliding motion parallel to the walls and as a consequence, the system with incommensurate packing reaches the arrested state at lower packing fractions, which facilitates reentrant glass transition along the constant packing fraction line. Furthermore, this study reveals a correlation between structural and dynamical evolution of confined liquids upon changing the wall separation, in particular we have shown that from the variation of the in-plane static structure one can anticipate a reentrant glass transition. These findings should be also of significant importance for all the cases where the microscopic behavior of confined liquids is required, for example, in biological and technological applications. Although the present study is restricted only to the long-time behavior of the MCT predictions, the full time-dependent solution of the intermediate scattering function remains to be tested and is the subject of future work.

The simulations presented here have been for Newtonian dynamics but we anticipate that the long-time dynamics also describes colloidal realizations of the confinement problem. For bulk systems comparisons between molecular dynamics or Brownian dynamics simulations and colloidal experiments suggest that the structural relaxation is independent of the microscopic dynamics, see Ref.~\cite{Pusey2009,Hunter2012} for recent reviews. For confined systems the hydrodynamic interactions will be different due to the interactions with the walls, nevertheless they are anticipated to induce only smooth changes of the dynamics which become negligible  in comparison to the singular behavior of the glassy dynamics.

In the present work we have considered paths in the nonequilibrium state diagram at constant packing fraction. A different choice would be to keep the chemical potential fixed  which corresponds to the situation of a wedge-like confinement of small opening angle where particle exchange along the wedge is permitted. It would also be interesting to compare the nonergodicity parameters of the MCT prediction to simulations or experiments following the glass transition line. However this will be extremely demanding both in computer simulations as well as in laboratory experiments.

\section{acknowledgments}
We thank Martin Oettel, Fathollah Varnik, and Rolf Schilling for useful discussions. The authors also acknowledge funding by Deutsche Forschungsgemeinschaft DFG via the research unit FOR1394 "Nonlinear Response to Probe Vitrification" and by the Austrian Science Fund (FWF): I 2887-N27.


\begin{thebibliography}{92}%
\makeatletter
\providecommand \@ifxundefined [1]{%
 \@ifx{#1\undefined}
}%
\providecommand \@ifnum [1]{%
 \ifnum #1\expandafter \@firstoftwo
 \else \expandafter \@secondoftwo
 \fi
}%
\providecommand \@ifx [1]{%
 \ifx #1\expandafter \@firstoftwo
 \else \expandafter \@secondoftwo
 \fi
}%
\providecommand \natexlab [1]{#1}%
\providecommand \enquote  [1]{``#1''}%
\providecommand \bibnamefont  [1]{#1}%
\providecommand \bibfnamefont [1]{#1}%
\providecommand \citenamefont [1]{#1}%
\providecommand \href@noop [0]{\@secondoftwo}%
\providecommand \href [0]{\begingroup \@sanitize@url \@href}%
\providecommand \@href[1]{\@@startlink{#1}\@@href}%
\providecommand \@@href[1]{\endgroup#1\@@endlink}%
\providecommand \@sanitize@url [0]{\catcode `\\12\catcode `\$12\catcode
  `\&12\catcode `\#12\catcode `\^12\catcode `\_12\catcode `\%12\relax}%
\providecommand \@@startlink[1]{}%
\providecommand \@@endlink[0]{}%
\providecommand \url  [0]{\begingroup\@sanitize@url \@url }%
\providecommand \@url [1]{\endgroup\@href {#1}{\urlprefix }}%
\providecommand \urlprefix  [0]{URL }%
\providecommand \Eprint [0]{\href }%
\providecommand \doibase [0]{http://dx.doi.org/}%
\providecommand \selectlanguage [0]{\@gobble}%
\providecommand \bibinfo  [0]{\@secondoftwo}%
\providecommand \bibfield  [0]{\@secondoftwo}%
\providecommand \translation [1]{[#1]}%
\providecommand \BibitemOpen [0]{}%
\providecommand \bibitemStop [0]{}%
\providecommand \bibitemNoStop [0]{.\EOS\space}%
\providecommand \EOS [0]{\spacefactor3000\relax}%
\providecommand \BibitemShut  [1]{\csname bibitem#1\endcsname}%
\let\auto@bib@innerbib\@empty
\bibitem [{\citenamefont {Adam}\ and\ \citenamefont {Gibbs}()}]{Adam1965}%
  \BibitemOpen
  \bibfield  {author} {\bibinfo {author} {\bibfnamefont {G.}~\bibnamefont
  {Adam}}\ and\ \bibinfo {author} {\bibfnamefont {J.~H.}\ \bibnamefont
  {Gibbs}},\ }\href {\doibase 10.1063/1.1696442} {\bibfield  {journal}
  {\bibinfo  {journal} {J. Chem. Phys.}\ }\textbf {\bibinfo {volume} {43}},\
  \bibinfo {pages} {139}}\BibitemShut {NoStop}%
\bibitem [{\citenamefont {Pusey}\ and\ \citenamefont {van
  Megen}(1987)}]{Pusey1987}%
  \BibitemOpen
  \bibfield  {author} {\bibinfo {author} {\bibfnamefont {P.~N.}\ \bibnamefont
  {Pusey}}\ and\ \bibinfo {author} {\bibfnamefont {W.}~\bibnamefont {van
  Megen}},\ }\href {\doibase 10.1103/PhysRevLett.59.2083} {\bibfield  {journal}
  {\bibinfo  {journal} {Phys. Rev. Lett.}\ }\textbf {\bibinfo {volume} {59}},\
  \bibinfo {pages} {2083} (\bibinfo {year} {1987})}\BibitemShut {NoStop}%
\bibitem [{\citenamefont {Stillinger}(1995)}]{Stillinger1995}%
  \BibitemOpen
  \bibfield  {author} {\bibinfo {author} {\bibfnamefont {F.~H.}\ \bibnamefont
  {Stillinger}},\ }\href {\doibase 10.1126/science.267.5206.1935} {\bibfield
  {journal} {\bibinfo  {journal} {Science}\ }\textbf {\bibinfo {volume}
  {267}},\ \bibinfo {pages} {1935} (\bibinfo {year} {1995})}\BibitemShut
  {NoStop}%
\bibitem [{\citenamefont {Ediger}\ \emph {et~al.}(1996)\citenamefont {Ediger},
  \citenamefont {Angell},\ and\ \citenamefont {Nagel}}]{Ediger1996}%
  \BibitemOpen
  \bibfield  {author} {\bibinfo {author} {\bibfnamefont {M.~D.}\ \bibnamefont
  {Ediger}}, \bibinfo {author} {\bibfnamefont {C.~A.}\ \bibnamefont {Angell}},
  \ and\ \bibinfo {author} {\bibfnamefont {S.~R.}\ \bibnamefont {Nagel}},\
  }\href {\doibase 10.1021/jp953538d} {\bibfield  {journal} {\bibinfo
  {journal} {J. Phys. Chem.}\ }\textbf {\bibinfo {volume} {100}},\ \bibinfo
  {pages} {13200} (\bibinfo {year} {1996})}\BibitemShut {NoStop}%
\bibitem [{\citenamefont {Debenedetti}\ and\ \citenamefont
  {Stillinger}(2001)}]{Debenedetti2001}%
  \BibitemOpen
  \bibfield  {author} {\bibinfo {author} {\bibfnamefont {P.~G.}\ \bibnamefont
  {Debenedetti}}\ and\ \bibinfo {author} {\bibfnamefont {F.~H.}\ \bibnamefont
  {Stillinger}},\ }\href {\doibase 10.1038/35065704} {\bibfield  {journal}
  {\bibinfo  {journal} {Nature}\ }\textbf {\bibinfo {volume} {410}},\ \bibinfo
  {pages} {259} (\bibinfo {year} {2001})}\BibitemShut {NoStop}%
\bibitem [{\citenamefont {Cipelletti}\ and\ \citenamefont
  {Ramos}(2005)}]{Cipelletti2005}%
  \BibitemOpen
  \bibfield  {author} {\bibinfo {author} {\bibfnamefont {L.}~\bibnamefont
  {Cipelletti}}\ and\ \bibinfo {author} {\bibfnamefont {L.}~\bibnamefont
  {Ramos}},\ }\href {http://stacks.iop.org/0953-8984/17/i=6/a=R01} {\bibfield
  {journal} {\bibinfo  {journal} {J. Phys. Condens. Matter}\ }\textbf {\bibinfo
  {volume} {17}},\ \bibinfo {pages} {R253} (\bibinfo {year}
  {2005})}\BibitemShut {NoStop}%
\bibitem [{\citenamefont {Biroli}\ \emph {et~al.}(2006)\citenamefont {Biroli},
  \citenamefont {Bouchaud}, \citenamefont {Miyazaki},\ and\ \citenamefont
  {Reichman}}]{Biroli2006}%
  \BibitemOpen
  \bibfield  {author} {\bibinfo {author} {\bibfnamefont {G.}~\bibnamefont
  {Biroli}}, \bibinfo {author} {\bibfnamefont {J.-P.}\ \bibnamefont
  {Bouchaud}}, \bibinfo {author} {\bibfnamefont {K.}~\bibnamefont {Miyazaki}},
  \ and\ \bibinfo {author} {\bibfnamefont {D.~R.}\ \bibnamefont {Reichman}},\
  }\href {\doibase 10.1103/PhysRevLett.97.195701} {\bibfield  {journal}
  {\bibinfo  {journal} {Phys. Rev. Lett.}\ }\textbf {\bibinfo {volume} {97}},\
  \bibinfo {pages} {195701} (\bibinfo {year} {2006})}\BibitemShut {NoStop}%
\bibitem [{\citenamefont {Heuer}(2008)}]{Heuer2008}%
  \BibitemOpen
  \bibfield  {author} {\bibinfo {author} {\bibfnamefont {A.}~\bibnamefont
  {Heuer}},\ }\href {\doibase :10.1088/0953-8984/20/37/373101} {\bibfield
  {journal} {\bibinfo  {journal} {J. Phys. Condens. Matter}\ }\textbf {\bibinfo
  {volume} {20}},\ \bibinfo {pages} {373101} (\bibinfo {year}
  {2008})}\BibitemShut {NoStop}%
\bibitem [{\citenamefont {Candelier}\ \emph {et~al.}(2010)\citenamefont
  {Candelier}, \citenamefont {Widmer-Cooper}, \citenamefont {Kummerfeld},
  \citenamefont {Dauchot}, \citenamefont {Biroli}, \citenamefont {Harrowell},\
  and\ \citenamefont {Reichman}}]{Candelier2010}%
  \BibitemOpen
  \bibfield  {author} {\bibinfo {author} {\bibfnamefont {R.}~\bibnamefont
  {Candelier}}, \bibinfo {author} {\bibfnamefont {A.}~\bibnamefont
  {Widmer-Cooper}}, \bibinfo {author} {\bibfnamefont {J.~K.}\ \bibnamefont
  {Kummerfeld}}, \bibinfo {author} {\bibfnamefont {O.}~\bibnamefont {Dauchot}},
  \bibinfo {author} {\bibfnamefont {G.}~\bibnamefont {Biroli}}, \bibinfo
  {author} {\bibfnamefont {P.}~\bibnamefont {Harrowell}}, \ and\ \bibinfo
  {author} {\bibfnamefont {D.~R.}\ \bibnamefont {Reichman}},\ }\href {\doibase
  10.1103/PhysRevLett.105.135702} {\bibfield  {journal} {\bibinfo  {journal}
  {Phys. Rev. Lett.}\ }\textbf {\bibinfo {volume} {105}},\ \bibinfo {pages}
  {135702} (\bibinfo {year} {2010})}\BibitemShut {NoStop}%
\bibitem [{\citenamefont {Berthier}(2011)}]{Berthier2011a}%
  \BibitemOpen
  \bibfield  {author} {\bibinfo {author} {\bibfnamefont {L.}~\bibnamefont
  {Berthier}},\ }\href {\doibase 10.1103/Physics.4.42} {\bibfield  {journal}
  {\bibinfo  {journal} {Physics}\ }\textbf {\bibinfo {volume} {4}},\ \bibinfo
  {pages} {42} (\bibinfo {year} {2011})}\BibitemShut {NoStop}%
\bibitem [{\citenamefont {Berthier}\ and\ \citenamefont
  {Biroli}(2011)}]{Berthier2011b}%
  \BibitemOpen
  \bibfield  {author} {\bibinfo {author} {\bibfnamefont {L.}~\bibnamefont
  {Berthier}}\ and\ \bibinfo {author} {\bibfnamefont {G.}~\bibnamefont
  {Biroli}},\ }\href {\doibase 10.1103/RevModPhys.83.587} {\bibfield  {journal}
  {\bibinfo  {journal} {Rev. Mod. Phys.}\ }\textbf {\bibinfo {volume} {83}},\
  \bibinfo {pages} {587} (\bibinfo {year} {2011})}\BibitemShut {NoStop}%
\bibitem [{\citenamefont {Hunter}\ and\ \citenamefont
  {Weeks}(2012)}]{Hunter2012}%
  \BibitemOpen
  \bibfield  {author} {\bibinfo {author} {\bibfnamefont {G.~L.}\ \bibnamefont
  {Hunter}}\ and\ \bibinfo {author} {\bibfnamefont {E.~R.}\ \bibnamefont
  {Weeks}},\ }\href {http://stacks.iop.org/0034-4885/75/i=6/a=066501}
  {\bibfield  {journal} {\bibinfo  {journal} {Rep. Prog. Phys.}\ }\textbf
  {\bibinfo {volume} {75}},\ \bibinfo {pages} {066501} (\bibinfo {year}
  {2012})}\BibitemShut {NoStop}%
\bibitem [{\citenamefont {Sengupta}\ \emph {et~al.}(2012)\citenamefont
  {Sengupta}, \citenamefont {Karmakar}, \citenamefont {Dasgupta},\ and\
  \citenamefont {Sastry}}]{Sengupta2012}%
  \BibitemOpen
  \bibfield  {author} {\bibinfo {author} {\bibfnamefont {S.}~\bibnamefont
  {Sengupta}}, \bibinfo {author} {\bibfnamefont {S.}~\bibnamefont {Karmakar}},
  \bibinfo {author} {\bibfnamefont {C.}~\bibnamefont {Dasgupta}}, \ and\
  \bibinfo {author} {\bibfnamefont {S.}~\bibnamefont {Sastry}},\ }\href
  {\doibase 10.1103/PhysRevLett.109.095705} {\bibfield  {journal} {\bibinfo
  {journal} {Phys. Rev. Lett.}\ }\textbf {\bibinfo {volume} {109}},\ \bibinfo
  {pages} {095705} (\bibinfo {year} {2012})}\BibitemShut {NoStop}%
\bibitem [{\citenamefont {Martinez-Garcia}\ \emph {et~al.}(2013)\citenamefont
  {Martinez-Garcia}, \citenamefont {Rzoska}, \citenamefont {Drozd-Rzoska},\
  and\ \citenamefont {Martinez-Garcia}}]{Martinez2013}%
  \BibitemOpen
  \bibfield  {author} {\bibinfo {author} {\bibfnamefont {J.~C.}\ \bibnamefont
  {Martinez-Garcia}}, \bibinfo {author} {\bibfnamefont {S.~J.}\ \bibnamefont
  {Rzoska}}, \bibinfo {author} {\bibfnamefont {A.}~\bibnamefont
  {Drozd-Rzoska}}, \ and\ \bibinfo {author} {\bibfnamefont {J.}~\bibnamefont
  {Martinez-Garcia}},\ }\href {\doibase 10.1038/ncomms2797} {\bibfield
  {journal} {\bibinfo  {journal} {Nat. Commun.}\ }\textbf {\bibinfo {volume}
  {4}},\ \bibinfo {pages} {1823} (\bibinfo {year} {2013})}\BibitemShut
  {NoStop}%
\bibitem [{\citenamefont {G\"otze}(2009)}]{Gotze_Complex_Dynamics}%
  \BibitemOpen
  \bibfield  {author} {\bibinfo {author} {\bibfnamefont {W.}~\bibnamefont
  {G\"otze}},\ }\href
  {http://www.oxfordscholarship.com/view/10.1093/acprof:oso/9780199235346.001.0001/acprof-9780199235346}
  {\emph {\bibinfo {title} {Complex Dynamics of Glass-Forming Liquids-A
  Mode-Coupling Theory}}}\ (\bibinfo  {publisher} {Oxford University},\
  \bibinfo {address} {Oxford},\ \bibinfo {year} {2009})\BibitemShut {NoStop}%
\bibitem [{\citenamefont {Voigtmann}\ and\ \citenamefont
  {Horbach}(2009)}]{Voigtmann2009}%
  \BibitemOpen
  \bibfield  {author} {\bibinfo {author} {\bibfnamefont {{\relax
  Th}.}~\bibnamefont {Voigtmann}}\ and\ \bibinfo {author} {\bibfnamefont
  {J.}~\bibnamefont {Horbach}},\ }\href {\doibase
  10.1103/PhysRevLett.103.205901} {\bibfield  {journal} {\bibinfo  {journal}
  {Phys. Rev. Lett.}\ }\textbf {\bibinfo {volume} {103}},\ \bibinfo {pages}
  {205901} (\bibinfo {year} {2009})}\BibitemShut {NoStop}%
\bibitem [{\citenamefont {Voigtmann}(2011)}]{Voigtmann2011}%
  \BibitemOpen
  \bibfield  {author} {\bibinfo {author} {\bibfnamefont {{\relax
  Th}.}~\bibnamefont {Voigtmann}},\ }\href
  {http://stacks.iop.org/0295-5075/96/i=3/a=36006} {\bibfield  {journal}
  {\bibinfo  {journal} {EPL}\ }\textbf {\bibinfo {volume} {96}},\ \bibinfo
  {pages} {36006} (\bibinfo {year} {2011})}\BibitemShut {NoStop}%
\bibitem [{\citenamefont {Sperl}\ \emph {et~al.}(2010)\citenamefont {Sperl},
  \citenamefont {Zaccarelli}, \citenamefont {Sciortino}, \citenamefont
  {Kumar},\ and\ \citenamefont {Stanley}}]{Sperl2010}%
  \BibitemOpen
  \bibfield  {author} {\bibinfo {author} {\bibfnamefont {M.}~\bibnamefont
  {Sperl}}, \bibinfo {author} {\bibfnamefont {E.}~\bibnamefont {Zaccarelli}},
  \bibinfo {author} {\bibfnamefont {F.}~\bibnamefont {Sciortino}}, \bibinfo
  {author} {\bibfnamefont {P.}~\bibnamefont {Kumar}}, \ and\ \bibinfo {author}
  {\bibfnamefont {H.~E.}\ \bibnamefont {Stanley}},\ }\href {\doibase
  10.1103/PhysRevLett.104.145701} {\bibfield  {journal} {\bibinfo  {journal}
  {Phys. Rev. Lett.}\ }\textbf {\bibinfo {volume} {104}},\ \bibinfo {pages}
  {145701} (\bibinfo {year} {2010})}\BibitemShut {NoStop}%
\bibitem [{\citenamefont {Gnan}\ \emph {et~al.}(2014)\citenamefont {Gnan},
  \citenamefont {Das}, \citenamefont {Sperl}, \citenamefont {Sciortino},\ and\
  \citenamefont {Zaccarelli}}]{Gnan2014}%
  \BibitemOpen
  \bibfield  {author} {\bibinfo {author} {\bibfnamefont {N.}~\bibnamefont
  {Gnan}}, \bibinfo {author} {\bibfnamefont {G.}~\bibnamefont {Das}}, \bibinfo
  {author} {\bibfnamefont {M.}~\bibnamefont {Sperl}}, \bibinfo {author}
  {\bibfnamefont {F.}~\bibnamefont {Sciortino}}, \ and\ \bibinfo {author}
  {\bibfnamefont {E.}~\bibnamefont {Zaccarelli}},\ }\href {\doibase
  10.1103/PhysRevLett.113.258302} {\bibfield  {journal} {\bibinfo  {journal}
  {Phys. Rev. Lett.}\ }\textbf {\bibinfo {volume} {113}},\ \bibinfo {pages}
  {258302} (\bibinfo {year} {2014})}\BibitemShut {NoStop}%
\bibitem [{\citenamefont {L\"owen}(2001)}]{Lowen2001}%
  \BibitemOpen
  \bibfield  {author} {\bibinfo {author} {\bibfnamefont {H.}~\bibnamefont
  {L\"owen}},\ }\href
  {http://iopscience.iop.org/article/10.1088/0953-8984/13/24/201/meta}
  {\bibfield  {journal} {\bibinfo  {journal} {J. Phys. Condens. Matter}\
  }\textbf {\bibinfo {volume} {13}},\ \bibinfo {pages} {R415} (\bibinfo {year}
  {2001})}\BibitemShut {NoStop}%
\bibitem [{\citenamefont {Alba-Simionesco}\ \emph {et~al.}(2006)\citenamefont
  {Alba-Simionesco}, \citenamefont {Coasne}, \citenamefont {Dosseh},
  \citenamefont {Dudziak}, \citenamefont {Gubbins}, \citenamefont
  {Radhakrishnan},\ and\ \citenamefont
  {Sliwinska-Bartkowiak}}]{Alba-Simionesco2006}%
  \BibitemOpen
  \bibfield  {author} {\bibinfo {author} {\bibfnamefont {C.}~\bibnamefont
  {Alba-Simionesco}}, \bibinfo {author} {\bibfnamefont {B.}~\bibnamefont
  {Coasne}}, \bibinfo {author} {\bibfnamefont {G.}~\bibnamefont {Dosseh}},
  \bibinfo {author} {\bibfnamefont {G.}~\bibnamefont {Dudziak}}, \bibinfo
  {author} {\bibfnamefont {K.~E.}\ \bibnamefont {Gubbins}}, \bibinfo {author}
  {\bibfnamefont {R.}~\bibnamefont {Radhakrishnan}}, \ and\ \bibinfo {author}
  {\bibfnamefont {M.}~\bibnamefont {Sliwinska-Bartkowiak}},\ }\href
  {http://iopscience.iop.org/article/10.1088/0953-8984/18/6/R01/meta}
  {\bibfield  {journal} {\bibinfo  {journal} {J. Phys. Condens. Matter}\
  }\textbf {\bibinfo {volume} {18}},\ \bibinfo {pages} {R15} (\bibinfo {year}
  {2006})}\BibitemShut {NoStop}%
\bibitem [{\citenamefont {Zhou}\ \emph {et~al.}(2008)\citenamefont {Zhou},
  \citenamefont {Rivas},\ and\ \citenamefont {Minton}}]{Zhou2008}%
  \BibitemOpen
  \bibfield  {author} {\bibinfo {author} {\bibfnamefont {H.-X.}\ \bibnamefont
  {Zhou}}, \bibinfo {author} {\bibfnamefont {G.}~\bibnamefont {Rivas}}, \ and\
  \bibinfo {author} {\bibfnamefont {A.~P.}\ \bibnamefont {Minton}},\ }\href
  {\doibase 10.1146/annurev.biophys.37.032807.125817} {\bibfield  {journal}
  {\bibinfo  {journal} {Annu. Rev. Biophys.}\ }\textbf {\bibinfo {volume}
  {37}},\ \bibinfo {pages} {375} (\bibinfo {year} {2008})}\BibitemShut
  {NoStop}%
\bibitem [{\citenamefont {Mattsson}\ \emph {et~al.}(2009)\citenamefont
  {Mattsson}, \citenamefont {Wyss}, \citenamefont {{Fernandez-Nieves}},
  \citenamefont {Miyazaki}, \citenamefont {Hu}, \citenamefont {Reichman},\ and\
  \citenamefont {Weitz}}]{Mattsson2009}%
  \BibitemOpen
  \bibfield  {author} {\bibinfo {author} {\bibfnamefont {J.}~\bibnamefont
  {Mattsson}}, \bibinfo {author} {\bibfnamefont {H.~M.}\ \bibnamefont {Wyss}},
  \bibinfo {author} {\bibfnamefont {A.}~\bibnamefont {{Fernandez-Nieves}}},
  \bibinfo {author} {\bibfnamefont {K.}~\bibnamefont {Miyazaki}}, \bibinfo
  {author} {\bibfnamefont {Z.}~\bibnamefont {Hu}}, \bibinfo {author}
  {\bibfnamefont {D.~R.}\ \bibnamefont {Reichman}}, \ and\ \bibinfo {author}
  {\bibfnamefont {D.~A.}\ \bibnamefont {Weitz}},\ }\href
  {http://www.nature.com/nature/journal/v462/n7269/full/nature08457.html}
  {\bibfield  {journal} {\bibinfo  {journal} {Nature}\ }\textbf {\bibinfo
  {volume} {462}},\ \bibinfo {pages} {83} (\bibinfo {year} {2009})}\BibitemShut
  {NoStop}%
\bibitem [{\citenamefont {Scheidler}\ \emph
  {et~al.}(2000{\natexlab{a}})\citenamefont {Scheidler}, \citenamefont {Kob},\
  and\ \citenamefont {Binder}}]{Scheidler2000}%
  \BibitemOpen
  \bibfield  {author} {\bibinfo {author} {\bibfnamefont {P.}~\bibnamefont
  {Scheidler}}, \bibinfo {author} {\bibfnamefont {W.}~\bibnamefont {Kob}}, \
  and\ \bibinfo {author} {\bibfnamefont {K.}~\bibnamefont {Binder}},\
  }\href@noop {} {\bibfield  {journal} {\bibinfo  {journal} {J. Phys. IV}\
  }\textbf {\bibinfo {volume} {10}},\ \bibinfo {pages} {33} (\bibinfo {year}
  {2000}{\natexlab{a}})}\BibitemShut {NoStop}%
\bibitem [{\citenamefont {Scheidler}\ \emph
  {et~al.}(2000{\natexlab{b}})\citenamefont {Scheidler}, \citenamefont {Kob},\
  and\ \citenamefont {Binder}}]{Scheidler2000b}%
  \BibitemOpen
  \bibfield  {author} {\bibinfo {author} {\bibfnamefont {P.}~\bibnamefont
  {Scheidler}}, \bibinfo {author} {\bibfnamefont {W.}~\bibnamefont {Kob}}, \
  and\ \bibinfo {author} {\bibfnamefont {K.}~\bibnamefont {Binder}},\ }\href
  {http://iopscience.iop.org/article/10.1209/epl/i2000-00435-1/meta} {\bibfield
   {journal} {\bibinfo  {journal} {Europhys. Lett.}\ }\textbf {\bibinfo
  {volume} {52}},\ \bibinfo {pages} {277} (\bibinfo {year}
  {2000}{\natexlab{b}})}\BibitemShut {NoStop}%
\bibitem [{\citenamefont {Scheidler}\ \emph {et~al.}(2002)\citenamefont
  {Scheidler}, \citenamefont {Kob},\ and\ \citenamefont
  {Binder}}]{Scheidler2002}%
  \BibitemOpen
  \bibfield  {author} {\bibinfo {author} {\bibfnamefont {P.}~\bibnamefont
  {Scheidler}}, \bibinfo {author} {\bibfnamefont {W.}~\bibnamefont {Kob}}, \
  and\ \bibinfo {author} {\bibfnamefont {K.}~\bibnamefont {Binder}},\ }\href
  {http://iopscience.iop.org/article/10.1209/epl/i2002-00182-9/meta} {\bibfield
   {journal} {\bibinfo  {journal} {EPL}\ }\textbf {\bibinfo {volume} {59}},\
  \bibinfo {pages} {701} (\bibinfo {year} {2002})}\BibitemShut {NoStop}%
\bibitem [{\citenamefont {Scheidler}\ \emph {et~al.}(2004)\citenamefont
  {Scheidler}, \citenamefont {Kob},\ and\ \citenamefont
  {Binder}}]{Scheidler2004}%
  \BibitemOpen
  \bibfield  {author} {\bibinfo {author} {\bibfnamefont {P.}~\bibnamefont
  {Scheidler}}, \bibinfo {author} {\bibfnamefont {W.}~\bibnamefont {Kob}}, \
  and\ \bibinfo {author} {\bibfnamefont {K.}~\bibnamefont {Binder}},\ }\href
  {http://pubs.acs.org/doi/abs/10.1021/jp036593s} {\bibfield  {journal}
  {\bibinfo  {journal} {J. Phys. Chem. B}\ }\textbf {\bibinfo {volume} {108}},\
  \bibinfo {pages} {6673} (\bibinfo {year} {2004})}\BibitemShut {NoStop}%
\bibitem [{\citenamefont {Varnik}\ \emph {et~al.}(2002)\citenamefont {Varnik},
  \citenamefont {Baschnagel},\ and\ \citenamefont {Binder}}]{Varnik2002c}%
  \BibitemOpen
  \bibfield  {author} {\bibinfo {author} {\bibfnamefont {F.}~\bibnamefont
  {Varnik}}, \bibinfo {author} {\bibfnamefont {J.}~\bibnamefont {Baschnagel}},
  \ and\ \bibinfo {author} {\bibfnamefont {K.}~\bibnamefont {Binder}},\ }\href
  {http://journals.aps.org/pre/abstract/10.1103/PhysRevE.65.021507} {\bibfield
  {journal} {\bibinfo  {journal} {Phy. Rev. E}\ }\textbf {\bibinfo {volume}
  {65}},\ \bibinfo {pages} {021507} (\bibinfo {year} {2002})}\BibitemShut
  {NoStop}%
\bibitem [{\citenamefont {Varnik}\ and\ \citenamefont
  {Binder}(2002)}]{Varnik2002d}%
  \BibitemOpen
  \bibfield  {author} {\bibinfo {author} {\bibfnamefont {F.}~\bibnamefont
  {Varnik}}\ and\ \bibinfo {author} {\bibfnamefont {K.}~\bibnamefont
  {Binder}},\ }\href
  {http://scitation.aip.org/content/aip/journal/jcp/117/13/10.1063/1.1503770}
  {\bibfield  {journal} {\bibinfo  {journal} {J. Chem. Phys.}\ }\textbf
  {\bibinfo {volume} {117}},\ \bibinfo {pages} {6336} (\bibinfo {year}
  {2002})}\BibitemShut {NoStop}%
\bibitem [{\citenamefont {Torres}\ \emph {et~al.}(2000)\citenamefont {Torres},
  \citenamefont {Nealey},\ and\ \citenamefont {de~Pablo}}]{Torres2000}%
  \BibitemOpen
  \bibfield  {author} {\bibinfo {author} {\bibfnamefont {J.~A.}\ \bibnamefont
  {Torres}}, \bibinfo {author} {\bibfnamefont {P.~F.}\ \bibnamefont {Nealey}},
  \ and\ \bibinfo {author} {\bibfnamefont {J.~J.}\ \bibnamefont {de~Pablo}},\
  }\href {http://journals.aps.org/prl/abstract/10.1103/PhysRevLett.85.3221}
  {\bibfield  {journal} {\bibinfo  {journal} {Phys. Rev. Lett.}\ }\textbf
  {\bibinfo {volume} {85}},\ \bibinfo {pages} {3221} (\bibinfo {year}
  {2000})}\BibitemShut {NoStop}%
\bibitem [{\citenamefont {Baschnagel}\ and\ \citenamefont
  {Varnik}(2005)}]{Baschnagel2005}%
  \BibitemOpen
  \bibfield  {author} {\bibinfo {author} {\bibfnamefont {J.}~\bibnamefont
  {Baschnagel}}\ and\ \bibinfo {author} {\bibfnamefont {F.}~\bibnamefont
  {Varnik}},\ }\href
  {http://iopscience.iop.org/article/10.1088/0953-8984/17/32/R02/meta}
  {\bibfield  {journal} {\bibinfo  {journal} {J. Phys. Condens. Matter}\
  }\textbf {\bibinfo {volume} {17}},\ \bibinfo {pages} {R851} (\bibinfo {year}
  {2005})}\BibitemShut {NoStop}%
\bibitem [{\citenamefont {Mittal}\ \emph {et~al.}(2006)\citenamefont {Mittal},
  \citenamefont {Errington},\ and\ \citenamefont {Truskett}}]{Mittal2006}%
  \BibitemOpen
  \bibfield  {author} {\bibinfo {author} {\bibfnamefont {J.}~\bibnamefont
  {Mittal}}, \bibinfo {author} {\bibfnamefont {J.~R.}\ \bibnamefont
  {Errington}}, \ and\ \bibinfo {author} {\bibfnamefont {T.~M.}\ \bibnamefont
  {Truskett}},\ }\href {\doibase 10.1103/PhysRevLett.96.177804} {\bibfield
  {journal} {\bibinfo  {journal} {Phys. Rev. Lett.}\ }\textbf {\bibinfo
  {volume} {96}},\ \bibinfo {pages} {177804} (\bibinfo {year}
  {2006})}\BibitemShut {NoStop}%
\bibitem [{\citenamefont {Mittal}\ \emph {et~al.}(2007)\citenamefont {Mittal},
  \citenamefont {Shen}, \citenamefont {Errington},\ and\ \citenamefont
  {Truskett}}]{Mittal2007}%
  \BibitemOpen
  \bibfield  {author} {\bibinfo {author} {\bibfnamefont {J.}~\bibnamefont
  {Mittal}}, \bibinfo {author} {\bibfnamefont {V.~K.}\ \bibnamefont {Shen}},
  \bibinfo {author} {\bibfnamefont {J.~R.}\ \bibnamefont {Errington}}, \ and\
  \bibinfo {author} {\bibfnamefont {T.~M.}\ \bibnamefont {Truskett}},\ }\href
  {http://scitation.aip.org/content/aip/journal/jcp/127/15/10.1063/1.2795699}
  {\bibfield  {journal} {\bibinfo  {journal} {J. Chem. Phys.}\ }\textbf
  {\bibinfo {volume} {127}},\ \bibinfo {pages} {154513} (\bibinfo {year}
  {2007})}\BibitemShut {NoStop}%
\bibitem [{\citenamefont {Jeetain}\ \emph {et~al.}(2007)\citenamefont
  {Jeetain}, \citenamefont {Jeffrey},\ and\ \citenamefont
  {Thomas}}]{Mittal2007b}%
  \BibitemOpen
  \bibfield  {author} {\bibinfo {author} {\bibfnamefont {M.}~\bibnamefont
  {Jeetain}}, \bibinfo {author} {\bibfnamefont {R.~E.}\ \bibnamefont
  {Jeffrey}}, \ and\ \bibinfo {author} {\bibfnamefont {M.~T.}\ \bibnamefont
  {Thomas}},\ }\href {\doibase 10.1021/jp071369e} {\bibfield  {journal}
  {\bibinfo  {journal} {J. Phys. Chem. B}\ }\textbf {\bibinfo {volume} {111}},\
  \bibinfo {pages} {10054} (\bibinfo {year} {2007})}\BibitemShut {NoStop}%
\bibitem [{\citenamefont {Mittal}\ \emph {et~al.}(2008)\citenamefont {Mittal},
  \citenamefont {Truskett}, \citenamefont {Errington},\ and\ \citenamefont
  {Hummer}}]{Mittal2008}%
  \BibitemOpen
  \bibfield  {author} {\bibinfo {author} {\bibfnamefont {J.}~\bibnamefont
  {Mittal}}, \bibinfo {author} {\bibfnamefont {T.~M.}\ \bibnamefont
  {Truskett}}, \bibinfo {author} {\bibfnamefont {J.~R.}\ \bibnamefont
  {Errington}}, \ and\ \bibinfo {author} {\bibfnamefont {G.}~\bibnamefont
  {Hummer}},\ }\href {\doibase 10.1103/PhysRevLett.100.145901} {\bibfield
  {journal} {\bibinfo  {journal} {Phys. Rev. Lett.}\ }\textbf {\bibinfo
  {volume} {100}},\ \bibinfo {pages} {145901} (\bibinfo {year}
  {2008})}\BibitemShut {NoStop}%
\bibitem [{\citenamefont {Krishnan}\ and\ \citenamefont
  {Ayappa}(2003)}]{Krishnan2003}%
  \BibitemOpen
  \bibfield  {author} {\bibinfo {author} {\bibfnamefont {S.~H.}\ \bibnamefont
  {Krishnan}}\ and\ \bibinfo {author} {\bibfnamefont {K.~G.}\ \bibnamefont
  {Ayappa}},\ }\href@noop {} {\bibfield  {journal} {\bibinfo  {journal} {The
  Journal of Chemical Physics}\ }\textbf {\bibinfo {volume} {118}} (\bibinfo
  {year} {2003})}\BibitemShut {NoStop}%
\bibitem [{\citenamefont {Krishnan}\ and\ \citenamefont
  {Ayappa}(2012)}]{Krishnan2012}%
  \BibitemOpen
  \bibfield  {author} {\bibinfo {author} {\bibfnamefont {S.~H.}\ \bibnamefont
  {Krishnan}}\ and\ \bibinfo {author} {\bibfnamefont {K.~G.}\ \bibnamefont
  {Ayappa}},\ }\href {\doibase 10.1103/PhysRevE.86.011504} {\bibfield
  {journal} {\bibinfo  {journal} {Phys. Rev. E}\ }\textbf {\bibinfo {volume}
  {86}},\ \bibinfo {pages} {011504} (\bibinfo {year} {2012})}\BibitemShut
  {NoStop}%
\bibitem [{\citenamefont {Goel}\ \emph {et~al.}(2008)\citenamefont {Goel},
  \citenamefont {Krekelberg}, \citenamefont {Errington},\ and\ \citenamefont
  {Truskett}}]{Goel2008}%
  \BibitemOpen
  \bibfield  {author} {\bibinfo {author} {\bibfnamefont {G.}~\bibnamefont
  {Goel}}, \bibinfo {author} {\bibfnamefont {W.~P.}\ \bibnamefont
  {Krekelberg}}, \bibinfo {author} {\bibfnamefont {J.~R.}\ \bibnamefont
  {Errington}}, \ and\ \bibinfo {author} {\bibfnamefont {T.~M.}\ \bibnamefont
  {Truskett}},\ }\href {\doibase 10.1103/PhysRevLett.100.106001} {\bibfield
  {journal} {\bibinfo  {journal} {Phys. Rev. Lett.}\ }\textbf {\bibinfo
  {volume} {100}},\ \bibinfo {pages} {106001} (\bibinfo {year}
  {2008})}\BibitemShut {NoStop}%
\bibitem [{\citenamefont {Krekelberg}\ \emph {et~al.}(2011)\citenamefont
  {Krekelberg}, \citenamefont {Shen}, \citenamefont {Errington},\ and\
  \citenamefont {Truskett}}]{Krekelberg2011}%
  \BibitemOpen
  \bibfield  {author} {\bibinfo {author} {\bibfnamefont {W.~P.}\ \bibnamefont
  {Krekelberg}}, \bibinfo {author} {\bibfnamefont {V.~K.}\ \bibnamefont
  {Shen}}, \bibinfo {author} {\bibfnamefont {J.~R.}\ \bibnamefont {Errington}},
  \ and\ \bibinfo {author} {\bibfnamefont {T.~M.}\ \bibnamefont {Truskett}},\
  }\href
  {http://scitation.aip.org/content/aip/journal/jcp/135/15/10.1063/1.3651478}
  {\bibfield  {journal} {\bibinfo  {journal} {J. Chem. Phys.}\ }\textbf
  {\bibinfo {volume} {135}},\ \bibinfo {eid} {154502} (\bibinfo {year}
  {2011})}\BibitemShut {NoStop}%
\bibitem [{\citenamefont {Ingebrigtsen}\ \emph {et~al.}(2013)\citenamefont
  {Ingebrigtsen}, \citenamefont {Errington}, \citenamefont {Truskett},\ and\
  \citenamefont {Dyre}}]{Ingebrigtsen2013}%
  \BibitemOpen
  \bibfield  {author} {\bibinfo {author} {\bibfnamefont {T.~S.}\ \bibnamefont
  {Ingebrigtsen}}, \bibinfo {author} {\bibfnamefont {J.~R.}\ \bibnamefont
  {Errington}}, \bibinfo {author} {\bibfnamefont {T.~M.}\ \bibnamefont
  {Truskett}}, \ and\ \bibinfo {author} {\bibfnamefont {J.~C.}\ \bibnamefont
  {Dyre}},\ }\href {\doibase 10.1103/PhysRevLett.111.235901} {\bibfield
  {journal} {\bibinfo  {journal} {Phys. Rev. Lett.}\ }\textbf {\bibinfo
  {volume} {111}},\ \bibinfo {pages} {235901} (\bibinfo {year}
  {2013})}\BibitemShut {NoStop}%
\bibitem [{\citenamefont {Ingebrigtsen}\ and\ \citenamefont
  {Dyre}(2014)}]{Ingebrigtsen2014}%
  \BibitemOpen
  \bibfield  {author} {\bibinfo {author} {\bibfnamefont {T.~S.}\ \bibnamefont
  {Ingebrigtsen}}\ and\ \bibinfo {author} {\bibfnamefont {J.~C.}\ \bibnamefont
  {Dyre}},\ }\href {\doibase 10.1039/C3SM52441H} {\bibfield  {journal}
  {\bibinfo  {journal} {Soft Matter}\ }\textbf {\bibinfo {volume} {10}},\
  \bibinfo {pages} {4324} (\bibinfo {year} {2014})}\BibitemShut {NoStop}%
\bibitem [{\citenamefont {Saw}\ and\ \citenamefont {Dasgupta}(2016)}]{Saw2016}%
  \BibitemOpen
  \bibfield  {author} {\bibinfo {author} {\bibfnamefont {S.}~\bibnamefont
  {Saw}}\ and\ \bibinfo {author} {\bibfnamefont {C.}~\bibnamefont {Dasgupta}},\
  }\href {\doibase 10.1063/1.4959942} {\bibfield  {journal} {\bibinfo
  {journal} {J. Chem. Phys.}\ }\textbf {\bibinfo {volume} {145}},\ \bibinfo
  {pages} {054707} (\bibinfo {year} {2016})}\BibitemShut {NoStop}%
\bibitem [{\citenamefont {Nugent}\ \emph {et~al.}(2007)\citenamefont {Nugent},
  \citenamefont {Edmond}, \citenamefont {Patel},\ and\ \citenamefont
  {Weeks}}]{Nugent2007}%
  \BibitemOpen
  \bibfield  {author} {\bibinfo {author} {\bibfnamefont {C.~R.}\ \bibnamefont
  {Nugent}}, \bibinfo {author} {\bibfnamefont {K.~V.}\ \bibnamefont {Edmond}},
  \bibinfo {author} {\bibfnamefont {H.~N.}\ \bibnamefont {Patel}}, \ and\
  \bibinfo {author} {\bibfnamefont {E.~R.}\ \bibnamefont {Weeks}},\ }\href
  {http://journals.aps.org/prl/abstract/10.1103/PhysRevLett.99.025702}
  {\bibfield  {journal} {\bibinfo  {journal} {Phys. Rev. Lett.}\ }\textbf
  {\bibinfo {volume} {99}},\ \bibinfo {pages} {025702} (\bibinfo {year}
  {2007})}\BibitemShut {NoStop}%
\bibitem [{\citenamefont {Sarangapani}\ \emph {et~al.}(2011)\citenamefont
  {Sarangapani}, \citenamefont {Schofield},\ and\ \citenamefont
  {Zhu}}]{Sarangapani2011}%
  \BibitemOpen
  \bibfield  {author} {\bibinfo {author} {\bibfnamefont {P.~S.}\ \bibnamefont
  {Sarangapani}}, \bibinfo {author} {\bibfnamefont {A.~B.}\ \bibnamefont
  {Schofield}}, \ and\ \bibinfo {author} {\bibfnamefont {Y.}~\bibnamefont
  {Zhu}},\ }\href {\doibase 10.1103/PhysRevE.83.030502} {\bibfield  {journal}
  {\bibinfo  {journal} {Phys. Rev. E}\ }\textbf {\bibinfo {volume} {83}},\
  \bibinfo {pages} {030502} (\bibinfo {year} {2011})}\BibitemShut {NoStop}%
\bibitem [{\citenamefont {Sarangapani}\ \emph {et~al.}(2012)\citenamefont
  {Sarangapani}, \citenamefont {Schofield},\ and\ \citenamefont
  {Zhu}}]{Sarangapani2012}%
  \BibitemOpen
  \bibfield  {author} {\bibinfo {author} {\bibfnamefont {P.~S.}\ \bibnamefont
  {Sarangapani}}, \bibinfo {author} {\bibfnamefont {A.~B.}\ \bibnamefont
  {Schofield}}, \ and\ \bibinfo {author} {\bibfnamefont {Y.}~\bibnamefont
  {Zhu}},\ }\href {\doibase 10.1039/C1SM06502E} {\bibfield  {journal} {\bibinfo
   {journal} {Soft Matter}\ }\textbf {\bibinfo {volume} {8}},\ \bibinfo {pages}
  {814} (\bibinfo {year} {2012})}\BibitemShut {NoStop}%
\bibitem [{\citenamefont {Hunter}\ \emph {et~al.}(2014)\citenamefont {Hunter},
  \citenamefont {Edmond},\ and\ \citenamefont {Weeks}}]{Hunter2014}%
  \BibitemOpen
  \bibfield  {author} {\bibinfo {author} {\bibfnamefont {G.~L.}\ \bibnamefont
  {Hunter}}, \bibinfo {author} {\bibfnamefont {K.~V.}\ \bibnamefont {Edmond}},
  \ and\ \bibinfo {author} {\bibfnamefont {E.~R.}\ \bibnamefont {Weeks}},\
  }\href {\doibase 10.1103/PhysRevLett.112.218302} {\bibfield  {journal}
  {\bibinfo  {journal} {Phys. Rev. Lett.}\ }\textbf {\bibinfo {volume} {112}},\
  \bibinfo {pages} {218302} (\bibinfo {year} {2014})}\BibitemShut {NoStop}%
\bibitem [{\citenamefont {Williams}\ \emph {et~al.}(2015)\citenamefont
  {Williams}, \citenamefont {O{\u{g}}uz}, \citenamefont {Bartlett},
  \citenamefont {L{\"o}wen},\ and\ \citenamefont
  {Patrick~Royall}}]{Williams2015}%
  \BibitemOpen
  \bibfield  {author} {\bibinfo {author} {\bibfnamefont {I.}~\bibnamefont
  {Williams}}, \bibinfo {author} {\bibfnamefont {E.~C.}\ \bibnamefont
  {O{\u{g}}uz}}, \bibinfo {author} {\bibfnamefont {P.}~\bibnamefont
  {Bartlett}}, \bibinfo {author} {\bibfnamefont {H.}~\bibnamefont {L{\"o}wen}},
  \ and\ \bibinfo {author} {\bibfnamefont {C.}~\bibnamefont {Patrick~Royall}},\
  }\href {\doibase 10.1063/1.4905472} {\bibfield  {journal} {\bibinfo
  {journal} {J. Chem. Phys.}\ }\textbf {\bibinfo {volume} {142}},\ \bibinfo
  {pages} {024505} (\bibinfo {year} {2015})}\BibitemShut {NoStop}%
\bibitem [{\citenamefont {Nyg\aa{}rd}\ \emph {et~al.}(2012)\citenamefont
  {Nyg\aa{}rd}, \citenamefont {Kjellander}, \citenamefont {Sarman},
  \citenamefont {Chodankar}, \citenamefont {Perret}, \citenamefont
  {Buitenhuis},\ and\ \citenamefont {van~der Veen}}]{Nygard2012}%
  \BibitemOpen
  \bibfield  {author} {\bibinfo {author} {\bibfnamefont {K.}~\bibnamefont
  {Nyg\aa{}rd}}, \bibinfo {author} {\bibfnamefont {R.}~\bibnamefont
  {Kjellander}}, \bibinfo {author} {\bibfnamefont {S.}~\bibnamefont {Sarman}},
  \bibinfo {author} {\bibfnamefont {S.}~\bibnamefont {Chodankar}}, \bibinfo
  {author} {\bibfnamefont {E.}~\bibnamefont {Perret}}, \bibinfo {author}
  {\bibfnamefont {J.}~\bibnamefont {Buitenhuis}}, \ and\ \bibinfo {author}
  {\bibfnamefont {J.~F.}\ \bibnamefont {van~der Veen}},\ }\href {\doibase
  10.1103/PhysRevLett.108.037802} {\bibfield  {journal} {\bibinfo  {journal}
  {Phys. Rev. Lett.}\ }\textbf {\bibinfo {volume} {108}},\ \bibinfo {pages}
  {037802} (\bibinfo {year} {2012})}\BibitemShut {NoStop}%
\bibitem [{\citenamefont {Nyg\aa{}rd}\ \emph {et~al.}(2013)\citenamefont
  {Nyg\aa{}rd}, \citenamefont {Sarman},\ and\ \citenamefont
  {Kjellander}}]{Nygard2013}%
  \BibitemOpen
  \bibfield  {author} {\bibinfo {author} {\bibfnamefont {K.}~\bibnamefont
  {Nyg\aa{}rd}}, \bibinfo {author} {\bibfnamefont {S.}~\bibnamefont {Sarman}},
  \ and\ \bibinfo {author} {\bibfnamefont {R.}~\bibnamefont {Kjellander}},\
  }\href
  {http://scitation.aip.org/content/aip/journal/jcp/139/16/10.1063/1.4825176}
  {\bibfield  {journal} {\bibinfo  {journal} {J. Chem. Phys.}\ }\textbf
  {\bibinfo {volume} {139}},\ \bibinfo {eid} {164701} (\bibinfo {year}
  {2013})}\BibitemShut {NoStop}%
\bibitem [{\citenamefont {Nyg\aa{}rd}\ \emph
  {et~al.}(2016{\natexlab{a}})\citenamefont {Nyg\aa{}rd}, \citenamefont
  {Sarman}, \citenamefont {Hyltegren}, \citenamefont {Chodankar}, \citenamefont
  {Perret}, \citenamefont {Buitenhuis}, \citenamefont {van~der Veen},\ and\
  \citenamefont {Kjellander}}]{Nygard2016a}%
  \BibitemOpen
  \bibfield  {author} {\bibinfo {author} {\bibfnamefont {K.}~\bibnamefont
  {Nyg\aa{}rd}}, \bibinfo {author} {\bibfnamefont {S.}~\bibnamefont {Sarman}},
  \bibinfo {author} {\bibfnamefont {K.}~\bibnamefont {Hyltegren}}, \bibinfo
  {author} {\bibfnamefont {S.}~\bibnamefont {Chodankar}}, \bibinfo {author}
  {\bibfnamefont {E.}~\bibnamefont {Perret}}, \bibinfo {author} {\bibfnamefont
  {J.}~\bibnamefont {Buitenhuis}}, \bibinfo {author} {\bibfnamefont {J.~F.}\
  \bibnamefont {van~der Veen}}, \ and\ \bibinfo {author} {\bibfnamefont
  {R.}~\bibnamefont {Kjellander}},\ }\href {\doibase 10.1103/PhysRevX.6.011014}
  {\bibfield  {journal} {\bibinfo  {journal} {Phys. Rev. X}\ }\textbf {\bibinfo
  {volume} {6}},\ \bibinfo {pages} {011014} (\bibinfo {year}
  {2016}{\natexlab{a}})}\BibitemShut {NoStop}%
\bibitem [{\citenamefont {Nyg\aa{}rd}\ \emph
  {et~al.}(2016{\natexlab{b}})\citenamefont {Nyg\aa{}rd}, \citenamefont
  {Buitenhuis}, \citenamefont {Kagias}, \citenamefont {Jefimovs}, \citenamefont
  {Zontone},\ and\ \citenamefont {Chushkin}}]{Nygard2016b}%
  \BibitemOpen
  \bibfield  {author} {\bibinfo {author} {\bibfnamefont {K.}~\bibnamefont
  {Nyg\aa{}rd}}, \bibinfo {author} {\bibfnamefont {J.}~\bibnamefont
  {Buitenhuis}}, \bibinfo {author} {\bibfnamefont {M.}~\bibnamefont {Kagias}},
  \bibinfo {author} {\bibfnamefont {K.}~\bibnamefont {Jefimovs}}, \bibinfo
  {author} {\bibfnamefont {F.}~\bibnamefont {Zontone}}, \ and\ \bibinfo
  {author} {\bibfnamefont {Y.}~\bibnamefont {Chushkin}},\ }\href {\doibase
  10.1103/PhysRevLett.116.167801} {\bibfield  {journal} {\bibinfo  {journal}
  {Phys. Rev. Lett.}\ }\textbf {\bibinfo {volume} {116}},\ \bibinfo {pages}
  {167801} (\bibinfo {year} {2016}{\natexlab{b}})}\BibitemShut {NoStop}%
\bibitem [{\citenamefont {Kienle}\ and\ \citenamefont
  {Kuhl}(2016)}]{Kienle2016}%
  \BibitemOpen
  \bibfield  {author} {\bibinfo {author} {\bibfnamefont {D.~F.}\ \bibnamefont
  {Kienle}}\ and\ \bibinfo {author} {\bibfnamefont {T.~L.}\ \bibnamefont
  {Kuhl}},\ }\href {\doibase 10.1103/PhysRevLett.117.036101} {\bibfield
  {journal} {\bibinfo  {journal} {Phys. Rev. Lett.}\ }\textbf {\bibinfo
  {volume} {117}},\ \bibinfo {pages} {036101} (\bibinfo {year}
  {2016})}\BibitemShut {NoStop}%
\bibitem [{\citenamefont {Zhang}\ and\ \citenamefont
  {Cheng}(2016)}]{Zhang2016}%
  \BibitemOpen
  \bibfield  {author} {\bibinfo {author} {\bibfnamefont {B.}~\bibnamefont
  {Zhang}}\ and\ \bibinfo {author} {\bibfnamefont {X.}~\bibnamefont {Cheng}},\
  }\href {\doibase 10.1103/PhysRevLett.116.098302} {\bibfield  {journal}
  {\bibinfo  {journal} {Phys. Rev. Lett.}\ }\textbf {\bibinfo {volume} {116}},\
  \bibinfo {pages} {098302} (\bibinfo {year} {2016})}\BibitemShut {NoStop}%
\bibitem [{\citenamefont {Ghosh}\ \emph {et~al.}(2016)\citenamefont {Ghosh},
  \citenamefont {Wijnperl{\'e}}, \citenamefont {Mugele},\ and\ \citenamefont
  {Duits}}]{Ghosh2016}%
  \BibitemOpen
  \bibfield  {author} {\bibinfo {author} {\bibfnamefont {S.}~\bibnamefont
  {Ghosh}}, \bibinfo {author} {\bibfnamefont {D.}~\bibnamefont
  {Wijnperl{\'e}}}, \bibinfo {author} {\bibfnamefont {F.}~\bibnamefont
  {Mugele}}, \ and\ \bibinfo {author} {\bibfnamefont {M.}~\bibnamefont
  {Duits}},\ }\href {\doibase 10.1039/C5SM02581H} {\bibfield  {journal}
  {\bibinfo  {journal} {Soft matter}\ }\textbf {\bibinfo {volume} {12}},\
  \bibinfo {pages} {1621} (\bibinfo {year} {2016})}\BibitemShut {NoStop}%
\bibitem [{\citenamefont {Gallo}\ \emph {et~al.}(2000)\citenamefont {Gallo},
  \citenamefont {Rovere},\ and\ \citenamefont {Spohr}}]{Gallo2000a}%
  \BibitemOpen
  \bibfield  {author} {\bibinfo {author} {\bibfnamefont {P.}~\bibnamefont
  {Gallo}}, \bibinfo {author} {\bibfnamefont {M.}~\bibnamefont {Rovere}}, \
  and\ \bibinfo {author} {\bibfnamefont {E.}~\bibnamefont {Spohr}},\ }\href
  {\doibase 10.1103/PhysRevLett.85.4317} {\bibfield  {journal} {\bibinfo
  {journal} {Phys. Rev. Lett.}\ }\textbf {\bibinfo {volume} {85}},\ \bibinfo
  {pages} {4317} (\bibinfo {year} {2000})}\BibitemShut {NoStop}%
\bibitem [{\citenamefont {Gallo}\ \emph {et~al.}(2009)\citenamefont {Gallo},
  \citenamefont {Attili},\ and\ \citenamefont {Rovere}}]{Gallo2009}%
  \BibitemOpen
  \bibfield  {author} {\bibinfo {author} {\bibfnamefont {P.}~\bibnamefont
  {Gallo}}, \bibinfo {author} {\bibfnamefont {A.}~\bibnamefont {Attili}}, \
  and\ \bibinfo {author} {\bibfnamefont {M.}~\bibnamefont {Rovere}},\ }\href
  {\doibase 10.1103/PhysRevE.80.061502} {\bibfield  {journal} {\bibinfo
  {journal} {Phys. Rev. E}\ }\textbf {\bibinfo {volume} {80}},\ \bibinfo
  {pages} {061502} (\bibinfo {year} {2009})}\BibitemShut {NoStop}%
\bibitem [{\citenamefont {Gallo}\ \emph {et~al.}(2012)\citenamefont {Gallo},
  \citenamefont {Rovere},\ and\ \citenamefont {Chen}}]{Gallo2012}%
  \BibitemOpen
  \bibfield  {author} {\bibinfo {author} {\bibfnamefont {P.}~\bibnamefont
  {Gallo}}, \bibinfo {author} {\bibfnamefont {M.}~\bibnamefont {Rovere}}, \
  and\ \bibinfo {author} {\bibfnamefont {S.-H.}\ \bibnamefont {Chen}},\ }\href
  {http://stacks.iop.org/0953-8984/24/i=6/a=064109} {\bibfield  {journal}
  {\bibinfo  {journal} {J. Phys. Condens. Matter}\ }\textbf {\bibinfo {volume}
  {24}},\ \bibinfo {pages} {064109} (\bibinfo {year} {2012})}\BibitemShut
  {NoStop}%
\bibitem [{\citenamefont {Krakoviack}(2005)}]{Krakoviack2005}%
  \BibitemOpen
  \bibfield  {author} {\bibinfo {author} {\bibfnamefont {V.}~\bibnamefont
  {Krakoviack}},\ }\href@noop {} {\bibfield  {journal} {\bibinfo  {journal}
  {Phys. Rev. Lett.}\ }\textbf {\bibinfo {volume} {94}},\ \bibinfo {pages}
  {065703} (\bibinfo {year} {2005})}\BibitemShut {NoStop}%
\bibitem [{\citenamefont {Krakoviack}(2007)}]{Krakoviack2007}%
  \BibitemOpen
  \bibfield  {author} {\bibinfo {author} {\bibfnamefont {V.}~\bibnamefont
  {Krakoviack}},\ }\href {\doibase 10.1103/PhysRevE.75.031503} {\bibfield
  {journal} {\bibinfo  {journal} {Phys. Rev. E}\ }\textbf {\bibinfo {volume}
  {75}},\ \bibinfo {eid} {031503} (\bibinfo {year} {2007})}\BibitemShut
  {NoStop}%
\bibitem [{\citenamefont {Krakoviack}(2009)}]{Krakoviack2009}%
  \BibitemOpen
  \bibfield  {author} {\bibinfo {author} {\bibfnamefont {V.}~\bibnamefont
  {Krakoviack}},\ }\href {\doibase 10.1103/PhysRevE.79.061501} {\bibfield
  {journal} {\bibinfo  {journal} {Phys. Rev. E}\ }\textbf {\bibinfo {volume}
  {79}},\ \bibinfo {eid} {061501} (\bibinfo {year} {2009})}\BibitemShut
  {NoStop}%
\bibitem [{\citenamefont {Krakoviack}(2011)}]{Krakoviack2011}%
  \BibitemOpen
  \bibfield  {author} {\bibinfo {author} {\bibfnamefont {V.}~\bibnamefont
  {Krakoviack}},\ }\href {\doibase 10.1103/PhysRevE.84.050501} {\bibfield
  {journal} {\bibinfo  {journal} {Phys. Rev. E}\ }\textbf {\bibinfo {volume}
  {84}},\ \bibinfo {pages} {050501} (\bibinfo {year} {2011})}\BibitemShut
  {NoStop}%
\bibitem [{\citenamefont {Szamel}\ and\ \citenamefont
  {Flenner}(2013)}]{Szamel2013}%
  \BibitemOpen
  \bibfield  {author} {\bibinfo {author} {\bibfnamefont {G.}~\bibnamefont
  {Szamel}}\ and\ \bibinfo {author} {\bibfnamefont {E.}~\bibnamefont
  {Flenner}},\ }\href {http://stacks.iop.org/0295-5075/101/i=6/a=66005}
  {\bibfield  {journal} {\bibinfo  {journal} {EPL}\ }\textbf {\bibinfo {volume}
  {101}},\ \bibinfo {pages} {66005} (\bibinfo {year} {2013})}\BibitemShut
  {NoStop}%
\bibitem [{\citenamefont {Kurzidim}\ \emph {et~al.}(2009)\citenamefont
  {Kurzidim}, \citenamefont {Coslovich},\ and\ \citenamefont
  {Kahl}}]{Kurzidim2009}%
  \BibitemOpen
  \bibfield  {author} {\bibinfo {author} {\bibfnamefont {J.}~\bibnamefont
  {Kurzidim}}, \bibinfo {author} {\bibfnamefont {D.}~\bibnamefont {Coslovich}},
  \ and\ \bibinfo {author} {\bibfnamefont {G.}~\bibnamefont {Kahl}},\ }\href
  {\doibase 10.1103/PhysRevLett.103.138303} {\bibfield  {journal} {\bibinfo
  {journal} {Phys. Rev. Lett.}\ }\textbf {\bibinfo {volume} {103}},\ \bibinfo
  {pages} {138303} (\bibinfo {year} {2009})}\BibitemShut {NoStop}%
\bibitem [{\citenamefont {Kurzidim}\ \emph {et~al.}(2010)\citenamefont
  {Kurzidim}, \citenamefont {Coslovich},\ and\ \citenamefont
  {Kahl}}]{Kurzidim2010}%
  \BibitemOpen
  \bibfield  {author} {\bibinfo {author} {\bibfnamefont {J.}~\bibnamefont
  {Kurzidim}}, \bibinfo {author} {\bibfnamefont {D.}~\bibnamefont {Coslovich}},
  \ and\ \bibinfo {author} {\bibfnamefont {G.}~\bibnamefont {Kahl}},\ }\href
  {\doibase 10.1103/PhysRevE.82.041505} {\bibfield  {journal} {\bibinfo
  {journal} {Phys. Rev. E}\ }\textbf {\bibinfo {volume} {82}},\ \bibinfo
  {pages} {041505} (\bibinfo {year} {2010})}\BibitemShut {NoStop}%
\bibitem [{\citenamefont {Kurzidim}\ \emph {et~al.}(2011)\citenamefont
  {Kurzidim}, \citenamefont {Coslovich},\ and\ \citenamefont
  {Kahl}}]{Kurzidim2011}%
  \BibitemOpen
  \bibfield  {author} {\bibinfo {author} {\bibfnamefont {J.}~\bibnamefont
  {Kurzidim}}, \bibinfo {author} {\bibfnamefont {D.}~\bibnamefont {Coslovich}},
  \ and\ \bibinfo {author} {\bibfnamefont {G.}~\bibnamefont {Kahl}},\ }\href
  {http://stacks.iop.org/0953-8984/23/i=23/a=234122} {\bibfield  {journal}
  {\bibinfo  {journal} {J. Phys.: Condens. Matter}\ }\textbf {\bibinfo {volume}
  {23}},\ \bibinfo {pages} {234122} (\bibinfo {year} {2011})}\BibitemShut
  {NoStop}%
\bibitem [{\citenamefont {Kim}\ \emph {et~al.}(2009)\citenamefont {Kim},
  \citenamefont {Miyazaki},\ and\ \citenamefont {Saito}}]{Kim2009}%
  \BibitemOpen
  \bibfield  {author} {\bibinfo {author} {\bibfnamefont {K.}~\bibnamefont
  {Kim}}, \bibinfo {author} {\bibfnamefont {K.}~\bibnamefont {Miyazaki}}, \
  and\ \bibinfo {author} {\bibfnamefont {S.}~\bibnamefont {Saito}},\ }\href
  {\doibase 10.1209/0295-5075/88/36002} {\bibfield  {journal} {\bibinfo
  {journal} {EPL}\ }\textbf {\bibinfo {volume} {88}},\ \bibinfo {pages} {36002}
  (\bibinfo {year} {2009})}\BibitemShut {NoStop}%
\bibitem [{\citenamefont {Kim}\ \emph {et~al.}(2011)\citenamefont {Kim},
  \citenamefont {Miyazaki},\ and\ \citenamefont {Saito}}]{Kim2011}%
  \BibitemOpen
  \bibfield  {author} {\bibinfo {author} {\bibfnamefont {K.}~\bibnamefont
  {Kim}}, \bibinfo {author} {\bibfnamefont {K.}~\bibnamefont {Miyazaki}}, \
  and\ \bibinfo {author} {\bibfnamefont {S.}~\bibnamefont {Saito}},\ }\href
  {http://stacks.iop.org/0953-8984/23/i=23/a=234123} {\bibfield  {journal}
  {\bibinfo  {journal} {J. Phys. Condens. Matter}\ }\textbf {\bibinfo {volume}
  {23}},\ \bibinfo {pages} {234123} (\bibinfo {year} {2011})}\BibitemShut
  {NoStop}%
\bibitem [{\citenamefont {Lang}\ \emph {et~al.}(2010)\citenamefont {Lang},
  \citenamefont {Bo\ifmmode~\mbox{\c{t}}\else \c{t}\fi{}an}, \citenamefont
  {Oettel}, \citenamefont {Hajnal}, \citenamefont {Franosch},\ and\
  \citenamefont {Schilling}}]{Lang2010}%
  \BibitemOpen
  \bibfield  {author} {\bibinfo {author} {\bibfnamefont {S.}~\bibnamefont
  {Lang}}, \bibinfo {author} {\bibfnamefont {V.}~\bibnamefont
  {Bo\ifmmode~\mbox{\c{t}}\else \c{t}\fi{}an}}, \bibinfo {author}
  {\bibfnamefont {M.}~\bibnamefont {Oettel}}, \bibinfo {author} {\bibfnamefont
  {D.}~\bibnamefont {Hajnal}}, \bibinfo {author} {\bibfnamefont
  {T.}~\bibnamefont {Franosch}}, \ and\ \bibinfo {author} {\bibfnamefont
  {R.}~\bibnamefont {Schilling}},\ }\href {\doibase
  10.1103/PhysRevLett.105.125701} {\bibfield  {journal} {\bibinfo  {journal}
  {Phys. Rev. Lett.}\ }\textbf {\bibinfo {volume} {105}},\ \bibinfo {pages}
  {125701} (\bibinfo {year} {2010})}\BibitemShut {NoStop}%
\bibitem [{\citenamefont {Lang}\ \emph {et~al.}(2012)\citenamefont {Lang},
  \citenamefont {Schilling}, \citenamefont {Krakoviack},\ and\ \citenamefont
  {Franosch}}]{Lang2012}%
  \BibitemOpen
  \bibfield  {author} {\bibinfo {author} {\bibfnamefont {S.}~\bibnamefont
  {Lang}}, \bibinfo {author} {\bibfnamefont {R.}~\bibnamefont {Schilling}},
  \bibinfo {author} {\bibfnamefont {V.}~\bibnamefont {Krakoviack}}, \ and\
  \bibinfo {author} {\bibfnamefont {T.}~\bibnamefont {Franosch}},\ }\href
  {\doibase 10.1103/PhysRevE.86.021502} {\bibfield  {journal} {\bibinfo
  {journal} {Phys. Rev. E}\ }\textbf {\bibinfo {volume} {86}},\ \bibinfo
  {pages} {021502} (\bibinfo {year} {2012})}\BibitemShut {NoStop}%
\bibitem [{\citenamefont {Lang}\ \emph {et~al.}(2013)\citenamefont {Lang},
  \citenamefont {Schilling},\ and\ \citenamefont {Franosch}}]{Lang2013b}%
  \BibitemOpen
  \bibfield  {author} {\bibinfo {author} {\bibfnamefont {S.}~\bibnamefont
  {Lang}}, \bibinfo {author} {\bibfnamefont {R.}~\bibnamefont {Schilling}}, \
  and\ \bibinfo {author} {\bibfnamefont {T.}~\bibnamefont {Franosch}},\ }\href
  {http://stacks.iop.org/1742-5468/2013/i=12/a=P12007} {\bibfield  {journal}
  {\bibinfo  {journal} {J. Stat. Mech. Theor. Exp}\ }\textbf {\bibinfo {volume}
  {2013}},\ \bibinfo {pages} {P12007} (\bibinfo {year} {2013})}\BibitemShut
  {NoStop}%
\bibitem [{\citenamefont {Lang}\ \emph
  {et~al.}(2014{\natexlab{a}})\citenamefont {Lang}, \citenamefont {Schilling},\
  and\ \citenamefont {Franosch}}]{Lang2014b}%
  \BibitemOpen
  \bibfield  {author} {\bibinfo {author} {\bibfnamefont {S.}~\bibnamefont
  {Lang}}, \bibinfo {author} {\bibfnamefont {R.}~\bibnamefont {Schilling}}, \
  and\ \bibinfo {author} {\bibfnamefont {T.}~\bibnamefont {Franosch}},\ }\href
  {\doibase 10.1103/PhysRevE.90.062126} {\bibfield  {journal} {\bibinfo
  {journal} {Phys. Rev. E}\ }\textbf {\bibinfo {volume} {90}},\ \bibinfo
  {pages} {062126} (\bibinfo {year} {2014}{\natexlab{a}})}\BibitemShut
  {NoStop}%
\bibitem [{\citenamefont {Franosch}\ \emph {et~al.}(2012)\citenamefont
  {Franosch}, \citenamefont {Lang},\ and\ \citenamefont
  {Schilling}}]{Franosch2012}%
  \BibitemOpen
  \bibfield  {author} {\bibinfo {author} {\bibfnamefont {T.}~\bibnamefont
  {Franosch}}, \bibinfo {author} {\bibfnamefont {S.}~\bibnamefont {Lang}}, \
  and\ \bibinfo {author} {\bibfnamefont {R.}~\bibnamefont {Schilling}},\ }\href
  {\doibase 10.1103/PhysRevLett.109.240601} {\bibfield  {journal} {\bibinfo
  {journal} {Phys. Rev. Lett.}\ }\textbf {\bibinfo {volume} {109}},\ \bibinfo
  {pages} {240601} (\bibinfo {year} {2012})}\BibitemShut {NoStop}%
\bibitem [{\citenamefont {Lang}\ \emph
  {et~al.}(2014{\natexlab{b}})\citenamefont {Lang}, \citenamefont {Franosch},\
  and\ \citenamefont {Schilling}}]{Lang2014c}%
  \BibitemOpen
  \bibfield  {author} {\bibinfo {author} {\bibfnamefont {S.}~\bibnamefont
  {Lang}}, \bibinfo {author} {\bibfnamefont {T.}~\bibnamefont {Franosch}}, \
  and\ \bibinfo {author} {\bibfnamefont {R.}~\bibnamefont {Schilling}},\ }\href
  {\doibase http://dx.doi.org/10.1063/1.4867284} {\bibfield  {journal}
  {\bibinfo  {journal} {J. Chem. Phys}\ }\textbf {\bibinfo {volume} {140}},\
  \bibinfo {eid} {104506} (\bibinfo {year} {2014}{\natexlab{b}})}\BibitemShut
  {NoStop}%
\bibitem [{\citenamefont {Mandal}\ and\ \citenamefont
  {Franosch}(2017)}]{Mandal2017a}%
  \BibitemOpen
  \bibfield  {author} {\bibinfo {author} {\bibfnamefont {S.}~\bibnamefont
  {Mandal}}\ and\ \bibinfo {author} {\bibfnamefont {T.}~\bibnamefont
  {Franosch}},\ }\href {\doibase 10.1103/PhysRevLett.118.065901} {\bibfield
  {journal} {\bibinfo  {journal} {Phys. Rev. Lett.}\ }\textbf {\bibinfo
  {volume} {118}},\ \bibinfo {pages} {065901} (\bibinfo {year}
  {2017})}\BibitemShut {NoStop}%
\bibitem [{\citenamefont {Lang}\ and\ \citenamefont
  {Franosch}(2014)}]{Lang2014a}%
  \BibitemOpen
  \bibfield  {author} {\bibinfo {author} {\bibfnamefont {S.}~\bibnamefont
  {Lang}}\ and\ \bibinfo {author} {\bibfnamefont {T.}~\bibnamefont
  {Franosch}},\ }\href {\doibase 10.1103/PhysRevE.89.062122} {\bibfield
  {journal} {\bibinfo  {journal} {Phys. Rev. E}\ }\textbf {\bibinfo {volume}
  {89}},\ \bibinfo {pages} {062122} (\bibinfo {year} {2014})}\BibitemShut
  {NoStop}%
\bibitem [{\citenamefont {Alder}\ and\ \citenamefont
  {Wainwright}(1957)}]{Alder1957}%
  \BibitemOpen
  \bibfield  {author} {\bibinfo {author} {\bibfnamefont {B.}~\bibnamefont
  {Alder}}\ and\ \bibinfo {author} {\bibfnamefont {T.}~\bibnamefont
  {Wainwright}},\ }\href {\doibase 10.1063/1.1743957} {\bibfield  {journal}
  {\bibinfo  {journal} {J. Chem. Phys.}\ }\textbf {\bibinfo {volume} {27}},\
  \bibinfo {pages} {1208} (\bibinfo {year} {1957})}\BibitemShut {NoStop}%
\bibitem [{\citenamefont {Rapaport}(1980)}]{Rapaport1980}%
  \BibitemOpen
  \bibfield  {author} {\bibinfo {author} {\bibfnamefont {D.}~\bibnamefont
  {Rapaport}},\ }\href
  {http://www.sciencedirect.com/science/article/pii/0021999180901047?via%3Dihub}
  {\bibfield  {journal} {\bibinfo  {journal} {J. Comput. Phys}\ }\textbf
  {\bibinfo {volume} {34}},\ \bibinfo {pages} {184} (\bibinfo {year}
  {1980})}\BibitemShut {NoStop}%
\bibitem [{\citenamefont {Bannerman}\ \emph {et~al.}(2011)\citenamefont
  {Bannerman}, \citenamefont {Sargant},\ and\ \citenamefont
  {Lue}}]{Bannerman2011}%
  \BibitemOpen
  \bibfield  {author} {\bibinfo {author} {\bibfnamefont {M.~N.}\ \bibnamefont
  {Bannerman}}, \bibinfo {author} {\bibfnamefont {R.}~\bibnamefont {Sargant}},
  \ and\ \bibinfo {author} {\bibfnamefont {L.}~\bibnamefont {Lue}},\ }\href
  {\doibase 10.1002/jcc.21915} {\bibfield  {journal} {\bibinfo  {journal} {J.
  Comput. Chem}\ }\textbf {\bibinfo {volume} {32}},\ \bibinfo {pages} {3329}
  (\bibinfo {year} {2011})}\BibitemShut {NoStop}%
\bibitem [{\citenamefont {Mandal}\ \emph {et~al.}(2014)\citenamefont {Mandal},
  \citenamefont {Lang}, \citenamefont {Gross}, \citenamefont {Oettel},
  \citenamefont {Raabe}, \citenamefont {Franosch},\ and\ \citenamefont
  {Varnik}}]{Mandal2014b}%
  \BibitemOpen
  \bibfield  {author} {\bibinfo {author} {\bibfnamefont {S.}~\bibnamefont
  {Mandal}}, \bibinfo {author} {\bibfnamefont {S.}~\bibnamefont {Lang}},
  \bibinfo {author} {\bibfnamefont {M.}~\bibnamefont {Gross}}, \bibinfo
  {author} {\bibfnamefont {M.}~\bibnamefont {Oettel}}, \bibinfo {author}
  {\bibfnamefont {D.}~\bibnamefont {Raabe}}, \bibinfo {author} {\bibfnamefont
  {T.}~\bibnamefont {Franosch}}, \ and\ \bibinfo {author} {\bibfnamefont
  {F.}~\bibnamefont {Varnik}},\ }\href {\doibase 10.1038/ncomms5435} {\bibfield
   {journal} {\bibinfo  {journal} {Nat. Commun.}\ }\textbf {\bibinfo {volume}
  {5}},\ \bibinfo {pages} {4435} (\bibinfo {year} {2014})}\BibitemShut
  {NoStop}%
\bibitem [{\citenamefont {Varnik}\ and\ \citenamefont
  {Franosch}(2016)}]{Varnik2016}%
  \BibitemOpen
  \bibfield  {author} {\bibinfo {author} {\bibfnamefont {F.}~\bibnamefont
  {Varnik}}\ and\ \bibinfo {author} {\bibfnamefont {T.}~\bibnamefont
  {Franosch}},\ }\href
  {http://iopscience.iop.org/article/10.1088/0953-8984/28/13/133001/meta}
  {\bibfield  {journal} {\bibinfo  {journal} {J. Phys. Condens. Matter}\
  }\textbf {\bibinfo {volume} {28}},\ \bibinfo {pages} {133001} (\bibinfo
  {year} {2016})}\BibitemShut {NoStop}%
\bibitem [{\citenamefont {Roth}(2010)}]{Roth2010}%
  \BibitemOpen
  \bibfield  {author} {\bibinfo {author} {\bibfnamefont {R.}~\bibnamefont
  {Roth}},\ }\href {\doibase 10.1088/0953-8984/22/6/063102} {\bibfield
  {journal} {\bibinfo  {journal} {J. Phys. Condens. Matter}\ }\textbf {\bibinfo
  {volume} {22}},\ \bibinfo {pages} {063102} (\bibinfo {year}
  {2010})}\BibitemShut {NoStop}%
\bibitem [{\citenamefont {Hansen-Goos}\ and\ \citenamefont
  {Roth}(2006)}]{Hansen2006}%
  \BibitemOpen
  \bibfield  {author} {\bibinfo {author} {\bibfnamefont {H.}~\bibnamefont
  {Hansen-Goos}}\ and\ \bibinfo {author} {\bibfnamefont {R.}~\bibnamefont
  {Roth}},\ }\href
  {http://iopscience.iop.org/article/10.1088/0953-8984/18/37/002/meta}
  {\bibfield  {journal} {\bibinfo  {journal} {J. Phys. Condens. Matter}\
  }\textbf {\bibinfo {volume} {18}},\ \bibinfo {pages} {8413} (\bibinfo {year}
  {2006})}\BibitemShut {NoStop}%
\bibitem [{\citenamefont {Hansen}\ and\ \citenamefont
  {McDonald}(2006)}]{Hansen:Theory_of_Simple_Liquids}%
  \BibitemOpen
  \bibfield  {author} {\bibinfo {author} {\bibfnamefont {J.~P.}\ \bibnamefont
  {Hansen}}\ and\ \bibinfo {author} {\bibfnamefont {I.~R.}\ \bibnamefont
  {McDonald}},\ }\href
  {http://www.sciencedirect.com/science/book/9780123705358} {\emph {\bibinfo
  {title} {Theory of Simple Liquids}}}\ (\bibinfo  {publisher} {Academic
  Press},\ \bibinfo {year} {2006})\BibitemShut {NoStop}%
\bibitem [{\citenamefont
  {Henderson}(1992)}]{Henderson:Fundamentals_of_inhomogeneous_fluids}%
  \BibitemOpen
  \bibfield  {author} {\bibinfo {author} {\bibfnamefont {D.}~\bibnamefont
  {Henderson}},\ }\href
  {https://www.crcpress.com/Fundamentals-of-Inhomogeneous-Fluids/Henderson/p/book/9780824787110}
  {\emph {\bibinfo {title} {Fundamentals of inhomogeneous fluids}}}\ (\bibinfo
  {publisher} {Dekker},\ \bibinfo {address} {New York},\ \bibinfo {year}
  {1992})\BibitemShut {NoStop}%
\bibitem [{\citenamefont {Ram}(2014)}]{Ram2014}%
  \BibitemOpen
  \bibfield  {author} {\bibinfo {author} {\bibfnamefont {J.}~\bibnamefont
  {Ram}},\ }\href
  {http://www.sciencedirect.com/science/article/pii/S0370157314000076}
  {\bibfield  {journal} {\bibinfo  {journal} {Physics Reports}\ }\textbf
  {\bibinfo {volume} {538}},\ \bibinfo {pages} {121 } (\bibinfo {year}
  {2014})}\BibitemShut {NoStop}%
\bibitem [{\citenamefont {Franosch}(2014)}]{Franosch2014}%
  \BibitemOpen
  \bibfield  {author} {\bibinfo {author} {\bibfnamefont {T.}~\bibnamefont
  {Franosch}},\ }\href {http://stacks.iop.org/1751-8121/47/i=32/a=325004}
  {\bibfield  {journal} {\bibinfo  {journal} {J. Phys. A-Math. Theor.}\
  }\textbf {\bibinfo {volume} {47}},\ \bibinfo {pages} {325004} (\bibinfo
  {year} {2014})}\BibitemShut {NoStop}%
\bibitem [{\citenamefont {Franosch}\ and\ \citenamefont {$\text{Th.}$
  Voigtmann}(2002)}]{Franosch2002}%
  \BibitemOpen
  \bibfield  {author} {\bibinfo {author} {\bibfnamefont {T.}~\bibnamefont
  {Franosch}}\ and\ \bibinfo {author} {\bibnamefont {$\text{Th.}$ Voigtmann}},\
  }\href {\doibase 10.1023/A:1019991729106} {\bibfield  {journal} {\bibinfo
  {journal} {{J. Stat. Phys.}}\ }\textbf {\bibinfo {volume} {{109}}},\ \bibinfo
  {pages} {{237}} (\bibinfo {year} {2002})}\BibitemShut {NoStop}%
\bibitem [{\citenamefont {Williams}\ and\ \citenamefont
  {Watts}(1970)}]{Williams1970}%
  \BibitemOpen
  \bibfield  {author} {\bibinfo {author} {\bibfnamefont {G.}~\bibnamefont
  {Williams}}\ and\ \bibinfo {author} {\bibfnamefont {D.~C.}\ \bibnamefont
  {Watts}},\ }\href {\doibase 10.1039/TF9706600080} {\bibfield  {journal}
  {\bibinfo  {journal} {Trans. Faraday Soc.}\ }\textbf {\bibinfo {volume}
  {66}},\ \bibinfo {pages} {80} (\bibinfo {year} {1970})}\BibitemShut {NoStop}%
\bibitem [{\citenamefont {Weysser}\ \emph {et~al.}(2010)\citenamefont
  {Weysser}, \citenamefont {Puertas}, \citenamefont {Fuchs},\ and\
  \citenamefont {Voigtmann}}]{Weysser2010}%
  \BibitemOpen
  \bibfield  {author} {\bibinfo {author} {\bibfnamefont {F.}~\bibnamefont
  {Weysser}}, \bibinfo {author} {\bibfnamefont {A.~M.}\ \bibnamefont
  {Puertas}}, \bibinfo {author} {\bibfnamefont {M.}~\bibnamefont {Fuchs}}, \
  and\ \bibinfo {author} {\bibfnamefont {T.}~\bibnamefont {Voigtmann}},\ }\href
  {\doibase 10.1103/PhysRevE.82.011504} {\bibfield  {journal} {\bibinfo
  {journal} {Phys. Rev. E}\ }\textbf {\bibinfo {volume} {82}},\ \bibinfo
  {pages} {011504} (\bibinfo {year} {2010})}\BibitemShut {NoStop}%
\bibitem [{\citenamefont {Voigtmann}(2003)}]{Voigtmann2003p}%
  \BibitemOpen
  \bibfield  {author} {\bibinfo {author} {\bibfnamefont {{\relax
  Th}.}~\bibnamefont {Voigtmann}},\ }\emph {\bibinfo {title} {Mode Coupling
  Theory of the Glass Transition in Binary Mixtures}},\ \href
  {https://mediatum.ub.tum.de/doc/603008/603008.pdf} {Ph.D. thesis},\ \bibinfo
  {school} {Technische Universit{\"a}t M{\"u}nchen} (\bibinfo {year}
  {2003})\BibitemShut {NoStop}%
\bibitem [{\citenamefont {Franosch}\ \emph {et~al.}(1997)\citenamefont
  {Franosch}, \citenamefont {Fuchs}, \citenamefont {G\"otze}, \citenamefont
  {Mayr},\ and\ \citenamefont {Singh}}]{Franosch1997}%
  \BibitemOpen
  \bibfield  {author} {\bibinfo {author} {\bibfnamefont {T.}~\bibnamefont
  {Franosch}}, \bibinfo {author} {\bibfnamefont {M.}~\bibnamefont {Fuchs}},
  \bibinfo {author} {\bibfnamefont {W.}~\bibnamefont {G\"otze}}, \bibinfo
  {author} {\bibfnamefont {M.~R.}\ \bibnamefont {Mayr}}, \ and\ \bibinfo
  {author} {\bibfnamefont {A.~P.}\ \bibnamefont {Singh}},\ }\href {\doibase
  10.1103/PhysRevE.56.5659} {\bibfield  {journal} {\bibinfo  {journal} {Phys.
  Rev. E}\ }\textbf {\bibinfo {volume} {56}},\ \bibinfo {pages} {5659}
  (\bibinfo {year} {1997})}\BibitemShut {NoStop}%
\bibitem [{\citenamefont {Pusey}\ \emph {et~al.}(2009)\citenamefont {Pusey},
  \citenamefont {Zaccarelli}, \citenamefont {Valeriani}, \citenamefont {Sanz},
  \citenamefont {Poon},\ and\ \citenamefont {Cates}}]{Pusey2009}%
  \BibitemOpen
  \bibfield  {author} {\bibinfo {author} {\bibfnamefont {P.}~\bibnamefont
  {Pusey}}, \bibinfo {author} {\bibfnamefont {E.}~\bibnamefont {Zaccarelli}},
  \bibinfo {author} {\bibfnamefont {C.}~\bibnamefont {Valeriani}}, \bibinfo
  {author} {\bibfnamefont {E.}~\bibnamefont {Sanz}}, \bibinfo {author}
  {\bibfnamefont {W.~C.}\ \bibnamefont {Poon}}, \ and\ \bibinfo {author}
  {\bibfnamefont {M.~E.}\ \bibnamefont {Cates}},\ }\href {\doibase
  10.1098/rsta.2009.0181} {\bibfield  {journal} {\bibinfo  {journal} {Phil.
  Trans. R. Soc. A}\ }\textbf {\bibinfo {volume} {367}},\ \bibinfo {pages}
  {4993} (\bibinfo {year} {2009})}\BibitemShut {NoStop}%
\end{thebibliography}

%

\end{document}